\documentclass[12pt,english,floatfix,superscriptaddress,aps,prd,preprint,showkeys,nofootinbib]{revtex4}
\usepackage{amsmath}
\usepackage{amssymb}
\usepackage{amsbsy}
\usepackage{amsfonts}
\usepackage{amsopn}
\usepackage{amstext}
\usepackage{graphicx}
\usepackage[english]{babel}
\usepackage{color}
\usepackage{slashed}
\usepackage{esint}
\usepackage[dvips]{epsfig}
\usepackage[dvips]{graphicx}
\usepackage{float}
\usepackage{units}
\usepackage{textcomp}
\usepackage{wasysym}
\usepackage{hyperref}
\usepackage{slashed}

\newcommand{\e}{\textrm{e}}

\begin{document}
\title{Thick braneworld in $f(Q,\mathcal{T})$ gravity}

\author{Fernando M. Belchior}
\email{belchior@fisica.ufc.br}
\affiliation{Departamento de F\'isica, Universidade Federal do Cear\'a,\\ Campus do Pici, 60455-760, Fortaleza, Cear\'a, Brazil.}

\author{Roberto V. Maluf}
\email{r.v.maluf@fisica.ufc.br}
\affiliation{Departamento de F\'isica, Universidade Federal do Cear\'a,\\ Campus do Pici, 60455-760, Fortaleza, Cear\'a, Brazil.}

\author{Carlos Alberto S. Almeida}
\email{carlos@fisica.ufc.br}
\affiliation{Departamento de F\'isica, Universidade Federal do Cear\'a,\\ Campus do Pici, 60455-760, Fortaleza, Cear\'a, Brazil.}

\begin{abstract}
In this work, we study a codimension-one thick brane within the framework of $f(Q,\mathcal{T})$ modified symmetric teleparallel gravity, where $Q$ is the nonmetricity scalar and $\mathcal{T}$ is the trace of the energy-momentum tensor. Using two specific choices of the warp factor, we construct the complete brane system, including the scalar field solution and energy density. We show how nonmetricity influences the brane profile and analyze the stability of the braneworld scenarios under small tensor perturbations.
\end{abstract}

\maketitle

\section{Introduction}

Recent astronomical observations have encouraged researchers to develop viable extensions of general relativity \cite{Berti:2015itd,Ishak:2018his}. Such extensions address significant issues in theoretical physics, such as baryon asymmetry \cite{Farrar:1993hn}, dark matter \cite{Navarro:1995iw}, dark energy (responsible for the accelerated expansion of the universe) \cite{Copeland:2006wr}, and the hierarchy problem \cite{Arkani-Hamed:1998jmv}. The latter has drawn attention to higher-dimensional models, also known as braneworld scenarios. This research topic gained prominence in string theory, particularly after the seminal papers by Lisa Randall and Raman Sundrum in 1999 \cite{rs,rs2}. 

The Randall-Sundrum (RS) model describes our universe as a 3-brane embedded in a warped higher-dimensional spacetime (bulk). In this framework, gravity is the only interaction that propagates through the bulk, while all other interactions remain confined to the 3-brane \cite{Charmousis2001}. It is important to note that the RS brane has a thin profile, which raises challenges regarding the localization of Standard Model fields, such as gauge and fermion fields \cite{Dubovsky:2001pe,Ringeval:2001cq}.  

Braneworld models with thick profiles, first introduced by Martim Gremm \cite{Gremm1999}, represent direct extensions of the RS model. These models are constructed by considering gravity coupled to a single scalar field with an interaction term. This term induces spontaneous symmetry breaking, leading to the formation of a domain wall.  

 In this context, aside from modifying the scalar field dynamics, one can also consider modified gravity models. Among these, $f(R)$ and $f(R,\mathcal{T})$ gravities stand out \cite{Afonso:2007gc,Chen:2020zzs,Almeida:2023kfl}, where $R$ represents the curvature scalar and $\mathcal{T}$ the trace of the energy-momentum tensor. In addition to these models, there exist curvature-free theories based on non-Riemannian geometry, such as $f(T)$, $f(T,B)$, $f(T,\mathcal{T})$, and $f(Q)$ gravities \cite{Yang2012,Menezes,ftmimetic,Fu:2021rgu,Silva:2022pfd}, where $T$ is the torsion scalar, $B$ the boundary term, and $Q$ the nonmetricity scalar.  Here, $f(T)$, $f(T,B)$, and $f(T,\mathcal{T})$ gravities are modifications of the teleparallel equivalent of general relativity (TEGR) \cite{ftenergyconditions, Bahamonde2015, Bahamonde:2021gfp,Ferraro2011us,Tamanini2012,Bahamonde2016,Bahamonde2020a,Harko:2014aja}, while $f(Q)$ gravity is a modification of the symmetric teleparallel equivalent of general relativity (STEGR) \cite{BeltranJimenez:2019esp,BeltranJimenez:2017tkd,BeltranJimenez:2019tme,Bajardi:2020fxh,Capozziello:2022tvvi,Capozziello:2022wgl,Gadbail:2022jco,Enkhili:2024dil}. Another extension of STEGR that has gained increasing attention in the literature is $f(Q,\mathcal{T})$ gravity \cite{Xu:2019sbp,Yang:2021fjy,Najera:2021afa,Arora:2020tuk,Arora:2020met,Gadbail:2023klq,Gadbail:2021fjf,ZeeshanGul:2024rqu}.  

This paper studies a five-dimensional thick brane in $f(Q,\mathcal{T})$ gravity coupled to a single scalar field. By choosing two specific warp factors, we determine the complete brane system. Furthermore, we investigate its stability under small perturbations and the localization of massless modes.  

The outline of this work is as follows: In Section (\ref{s2}), we introduce the fundamental concepts of STERG and derive the equations of motion for $f(Q,\mathcal{T})$ gravity. In Section (\ref{s3}), we construct the thick brane scenario within $f(Q,\mathcal{T})$ gravity by considering two specific choices for the warp factor. In Section (\ref{s4}), we examine the stability of the brane system under small perturbations and analyze the localization of massless modes. Finally, in Section (\ref{s5}), we present our concluding remarks and discuss future perspectives. Throughout this paper, we adopt natural units and use a metric with signature $(-,+,+,+,+)$.  

\section{$f(Q,\mathcal{T})$ gravity and thick brane}
\label{s2}

As a starting point, let us introduce the general concepts of symmetric teleparallel gravity and present the equations of motion for $f(Q,\mathcal{T})$ gravity in this section. In theories based on Riemannian geometry, the metricity condition $\nabla_M g_{NP} = 0$ must be satisfied, where $g_{NP}$ denotes the metric and $\nabla_M$ is the covariant derivative associated with the Levi-Civita connection $\Gamma^{P}\ _{MN}$. We denote the bulk coordinate indices by capital Latin index $M=0,\ldots, D-1$. For example, general relativity (GR) and $f(R)$ gravity satisfy this condition. However, this relation is no longer satisfied in theories based on non-Riemannian geometry. Among these modified gravity models, symmetric teleparallel equivalent of general relativity (STEGR) is characterized by a nonvanishing nonmetricity tensor \cite{Nester:1998mp}:
\begin{align}
Q_{MNP}=\nabla_M g_{NP}.   
\end{align}
For this tensor, we define its independent traces as $Q_M = g^{NP} Q_{MNP}$ and $\widetilde{Q}_M = g^{NP} Q_{NMP}$. Additionally, due to the presence of the nonmetricity tensor, a more general connection $\widetilde{\Gamma}^P\ _{MN}$ must be introduced in the context of STEGR, given by  
\begin{align}
\widetilde{\Gamma}^P\ _{MN} = \Gamma^P\ _{MN} + L^P\ _{MN},    
\end{align}
where $L^P\ _{MN}$ is the distortion tensor, expressed in terms of the nonmetricity tensor as follows \cite{Nester:1998mp}:  
\begin{align}
L^P\ _{MN} = \frac{1}{2} g^{PQ} (Q_{PMN} - Q_{MPN} - Q_{NPM}).    
\end{align}

To construct a gravitational action for STEGR, we introduce a more general tensor that incorporates the nonmetricity, its independent traces, and the distortion tensor. This tensor is known as the nonmetricity conjugate \cite{Nester:1998mp}, and is explicitly given by  
\begin{align}
P^P\ _{MN} = -\frac{1}{2} L^P\ _{MN} + \frac{1}{4} (Q^P - \widetilde{Q}^P) g_{MN} - \frac{1}{8} (\delta^P_M Q_N + \delta^P_N Q_M).    
\end{align}
Furthermore, its contraction with the nonmetricity tensor provides the nonmetricity scalar $Q = Q_{PMN} P^{PMN}$. Additionally, we define the Ricci scalar as $R = Q + B$, where $B = \nabla_M (Q^M - \widetilde{Q}^M)$ is a boundary term. This result shows that STEGR is equivalent to GR, since the boundary term vanishes upon integration into the action. However, when dealing with $f(Q)$ and $f(R)$ gravities, the boundary term becomes relevant.  

In this work, we focus on $f(Q, \mathcal{T})$ gravity, which represents another possible extension of STEGR \cite{Xu:2019sbp, Yang:2021fjy, Najera:2021afa, Arora:2020tuk, Arora:2020met, Gadbail:2023klq, Gadbail:2021fjf}. The five-dimensional gravitational action is given by  
\begin{align}\label{a1}
S = \int d^5 x \sqrt{-g} \left[\frac{1}{4} f(Q, \mathcal{T}) + \mathcal{L}_m \right],  
\end{align}
where $\mathcal{L}_m$ is the matter Lagrangian, to be defined later. The variation of the action (\ref{a1}) with respect to the metric leads to the following modified Einstein equation:  
\begin{align}
G_{MN} = 2 \left[\mathcal{T}_{MN} - \frac{f_T}{2} (\mathcal{T}_{MN} + \theta_{MN}) \right],  
\end{align}where  
\begin{align}\label{graeq}
G_{MN} = \frac{2}{\sqrt{-g}} \nabla_K \left( \sqrt{-g} f_Q P^K\ _{MN} \right) - \frac{1}{2} g_{MN} f + f_Q \left( P_{MKL} Q_N\ ^{KL} - 2 Q^L\ _{KM} P^K\ _{NL} \right).  
\end{align}

In addition, $\mathcal{T}_{MN}$ is the energy-momentum tensor defined by
\begin{align}
\mathcal{T}_{MN}=-2\frac{\delta \mathcal{L}_m}{\delta g^{MN}}+ g_{MN}\mathcal{L}_m,
\end{align}
while we define the tensor $\theta_{MN}$ as follows
\begin{align}
\theta_{MN}=g^{AB}\frac{\delta T_{AB}}{\delta g^{MN}}.
\end{align}

At the same time, by varying the action (\ref{a1}) with respect to the connection, we obtain
\begin{align}\label{coneq}
\nabla_M \nabla_N \left( \sqrt{-g} f_Q P_K\ ^{MN} \right) = 0.   
\end{align}
For simplicity, we define $f \equiv f(Q,T)$, $f_Q \equiv \frac{\partial f(Q,T)}{\partial Q}$, and $f_T \equiv \frac{\partial f(Q,T)}{\partial T}$.

Once we have revised some concepts of STEGR, we are now able to 
build the braneworld scenario. In this sense, let us consider a single scalar field with standard dynamics and an interaction potential as the matter source. This scalar field depends only on an extra dimension and provides thickness for the brane. Its Lagrangian reads
\begin{align}
\mathcal{L}_m=-\frac{1}{2}\partial_M\phi\partial^M\phi-V(\phi).   
\end{align}
Concurrently, the energy-momentum tensor associated to this Lagrangian assumes the form
\begin{align}
\mathcal{T}_{MN}=\partial_M\phi\partial_N\phi+g_{MN}\mathcal{L}_m,    
\end{align}
while its trace reads
\begin{align}
\mathcal{T}=-\frac{3}{2}\partial_M\phi\partial^M\phi-5V. 
\end{align}
Thus, we write the tensor $\theta_{MN}$ as being 
\begin{align}
\theta_{MN}= -\frac{5}{2}\partial_M\phi\partial_N\phi-g_{MN}\mathcal{L}_m.   
\end{align}

In order to obtain the brane equations, we consider a flat Randall-Sundrum-like metric, namely
\begin{align}\label{metric}
ds^2= \e^{2A}\eta_{\mu\nu}dx^{\mu}dx^{\nu}+dy^2,    
\end{align}
where $\eta_{\mu\nu}$ is the Minkowski metric, $\e^{2A}$ is the warp factor and $y$ represent the extra dimension. Like the scalar field, $A(y)$ is only a function of the extra dimension. The greek indices $\mu,\nu$ run from 0 to 3. Besides, we should point out that we will choose the coincident gauge, i.e., $\widetilde{\Gamma}^P\ _{MN}=0$. Once we have chosen the coincident gauge, the Eq. (\ref{coneq}) is satisfied. Then, the scalar field and gravitational equations read
\begin{eqnarray}
\Big(1+\frac{3}{4}f_{\mathcal{T}}\Big)\phi^{\prime\prime}+\Big[(4+3f_{\mathcal{T}})A^{\prime}+\frac{3}{4} f_{\mathcal{T}}^{\prime}\Big]\phi^{\prime}&=&\Big(1+\frac{5}{4}f_{\mathcal{T}}\Big)V_{\phi},\label{erer}
\\
12f_Q A^{\prime 2}-\frac{1}{2}f&=&\Big(1+\frac{3}{2}f_{\mathcal{T}}\Big)\phi^{\prime 2}-2V,\label{erer2}
\\
3(f_Q A^{\prime\prime}+f_Q^{\prime}A^{\prime })&=&-\Big(2+\frac{3}{2}f_{\mathcal{T}}\Big)\phi^{\prime 2}\label{erer3}
.
\end{eqnarray}
In the above set of equations, the prime ($^{\prime}$) denotes derivative with respect to extra dimension. Besides, allow us assume a form for the function $f(Q,\mathcal{T})$, specifically
\begin{align}
f(Q,\mathcal{T})=Q+k_1\, Q^n+k_2\,\mathcal{T},    
\end{align}
where the parameters $k_1$ and $n$ refer to influence of nonmetricity, whereas the parameter $k_2$ is related to the trace of the energy-momentum tensor. 

The next step is to solve the set of equations (\ref{erer}-\ref{erer3}). For this purpose, we assume a suitable form for the warp factor. Thereby, to make a choice, we require the warp factor to obey some requirements listed below:
\begin{itemize}
 \item Far from brane core, we must have $\lim_{y\rightarrow \infty}\e^{2A(y)}\rightarrow 0$, reproducing the RS warp factor;
 \item Near the brane, the warp factor should have a smooth profile;
  \item To ensure that the matter field has $Z_2$ symmetry $\e^{2A(y)}=\e^{2A(-y)}$;
   \item $\int_{-\infty}^{\infty}dy \e^{8A(y)}$ must be finite and non-null to ensure a normalizable zero-mode for graviton.
\end{itemize}
Based on these requirements, we choose the following two forms for warp factor: 
\begin{align}
    &A(y)=\omega \log{[\mathrm{sech}(ty)}],\\
    &A(y)=\log{[\mathrm{tanh}[t(y+c)]-\mathrm{tanh}[t(y+c)]]}.
\end{align}
As will be discussed, the parameters $\omega$, $t$ and $c$ play a significant role in the brane system and the internal structure. Once a form for the warp factor is chosen, the next step is to solve the system of equations to find the matter field and the energy density, which has the analytical form
\begin{align}
\rho(y) = e^{2A(y)} \left( \frac{1}{2} \phi^{\prime\, 2} + V(\phi) \right).
\end{align}
In the following sections, we will examine each of the two warp factor forms separately.

\section{Brane system}\label{s3}

\subsection{Model 1}

\begin{figure}[ht!]
\begin{center}
\begin{tabular}{ccc}
\includegraphics[height=5cm]{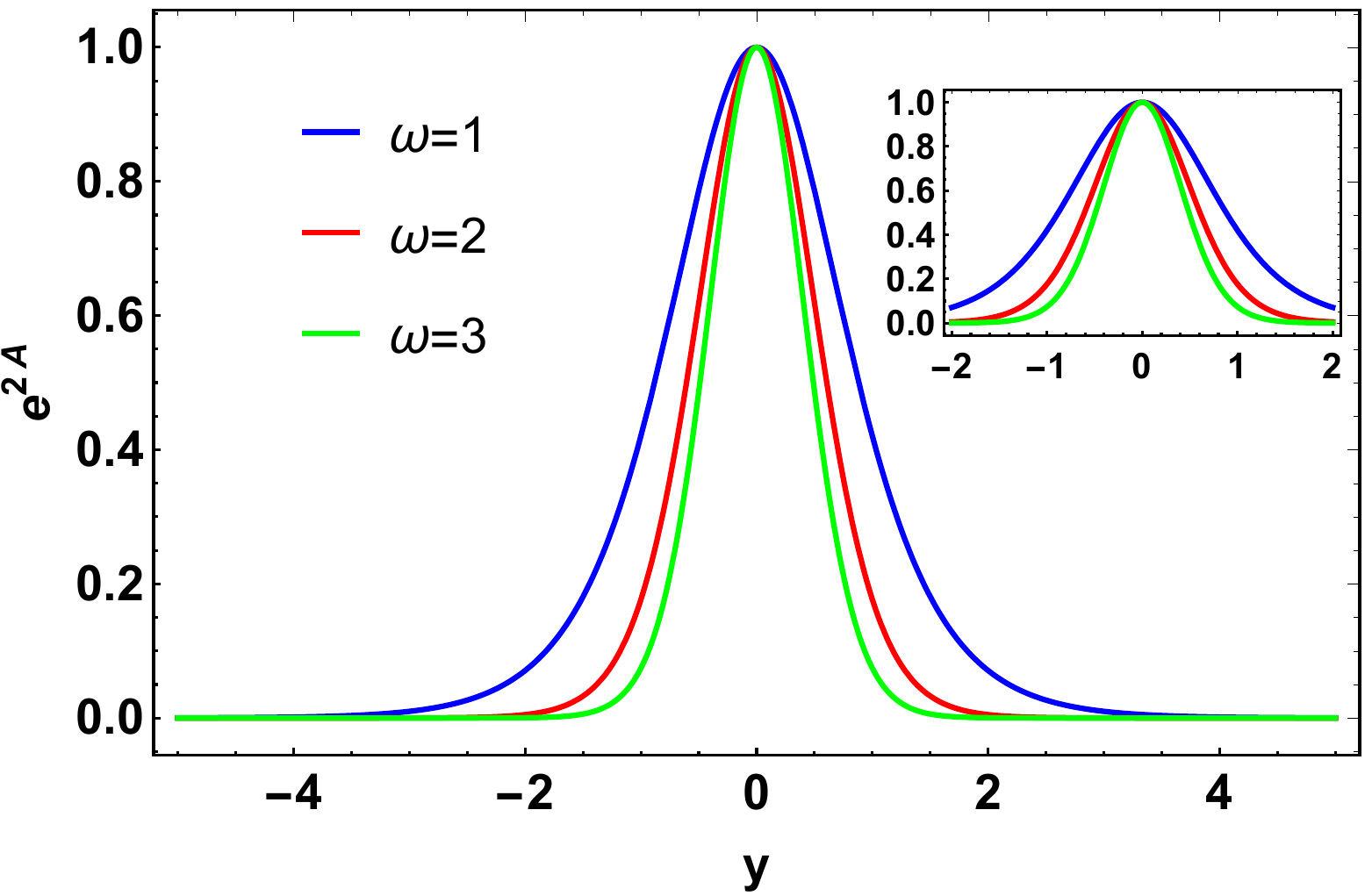} 
\includegraphics[height=5cm]{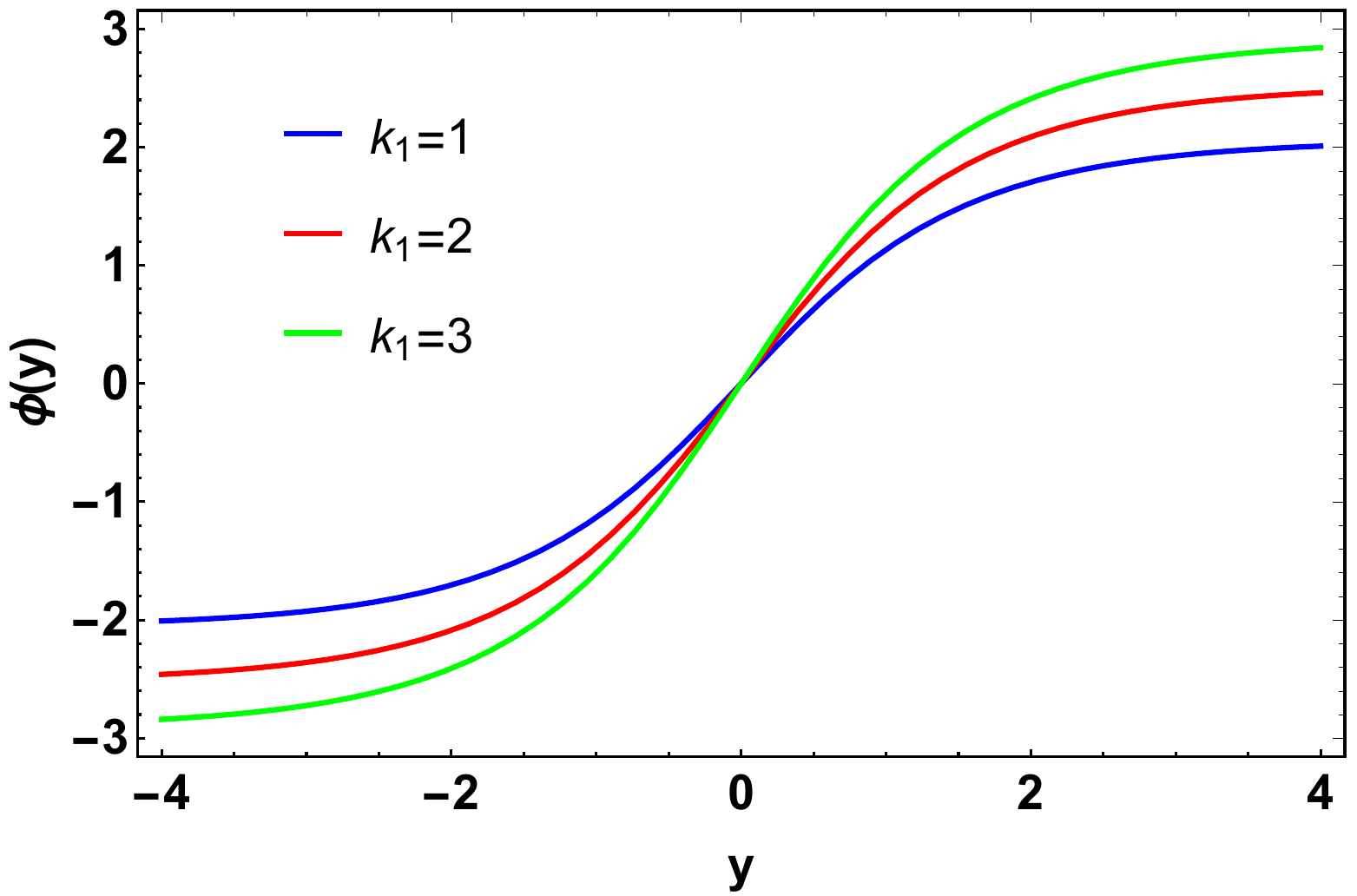}\\
(a)\hspace{6.7cm}(b)\\
\includegraphics[height=5cm]{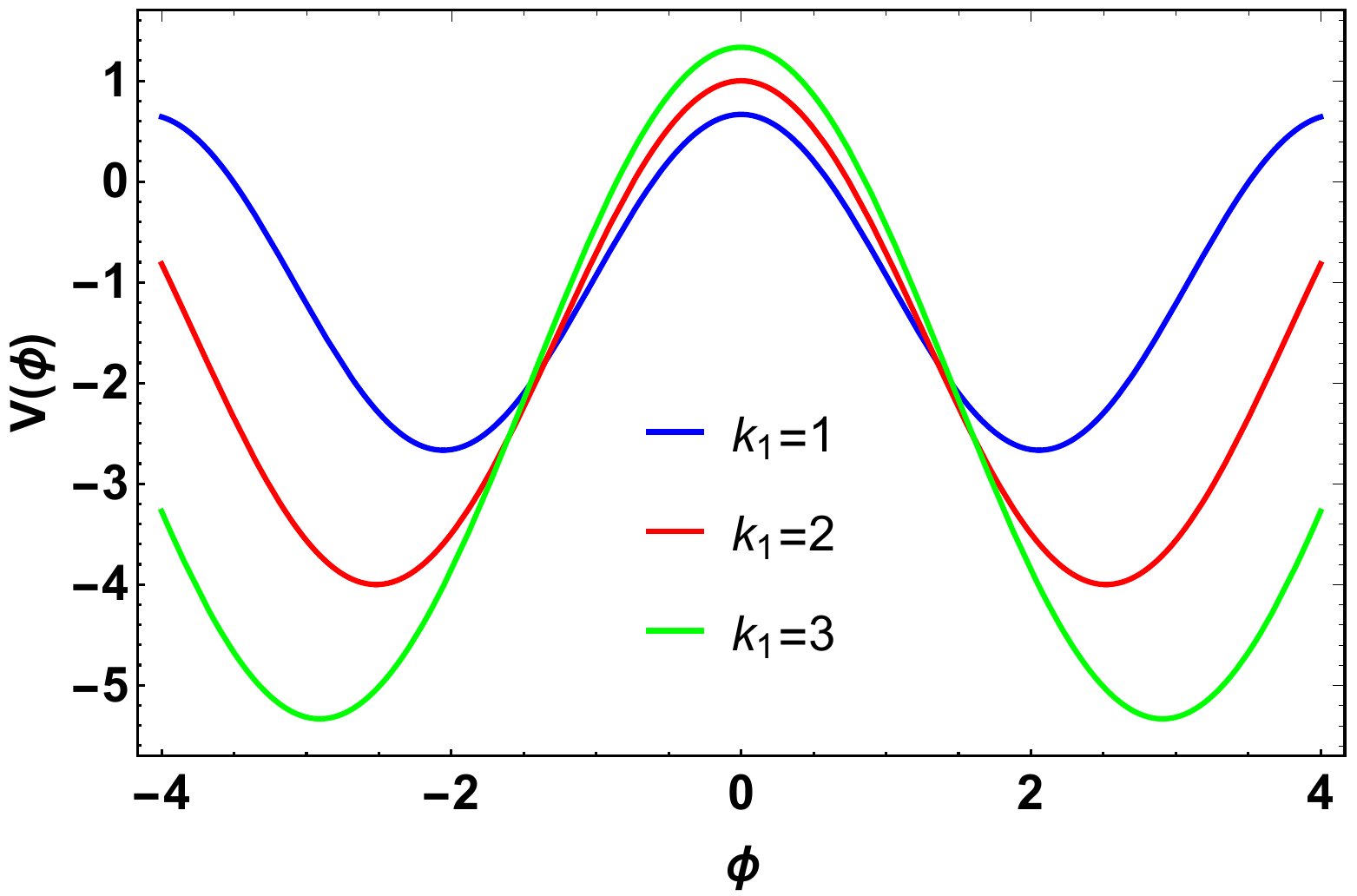}
\includegraphics[height=5cm]{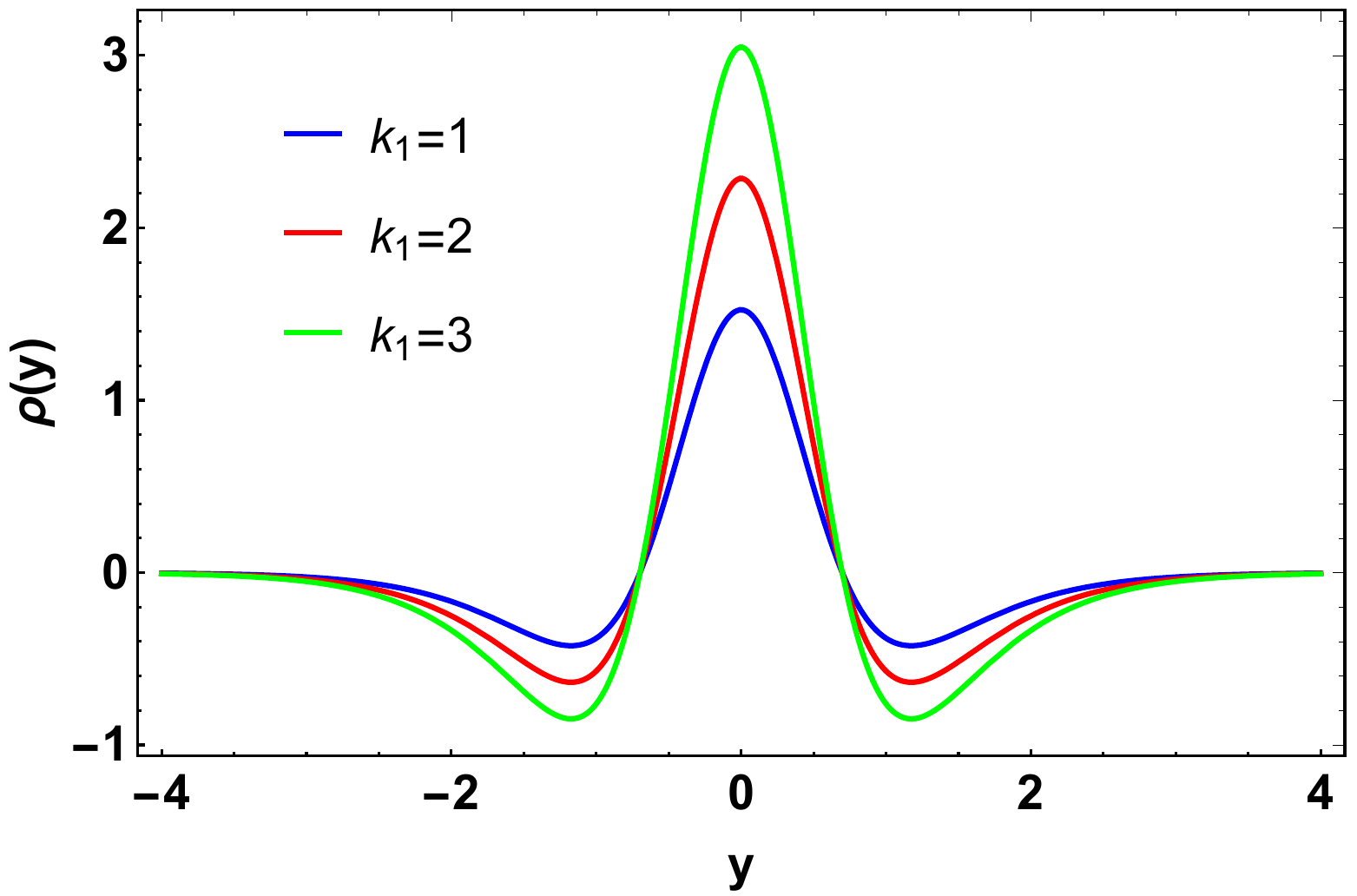}
\\
(c)\hspace{6.7cm}(d) 
\end{tabular}
\end{center}
\vspace{-0.5cm}
\caption{For model 1 with $n=1$. (a) Warp factor. (b) Scalar field. (c) Potential. (d) Energy density.
\label{fig1}}
\end{figure}

As first model, we consider the following warp factor 
\begin{align}\label{wf1}
    A(y)=\omega \log{[\mathrm{sech}(ty)}].
\end{align}
As depicted in Fig. (\ref{fig1})(a), the warp factor displays a smooth behavior near the brane core. For this warp factor, the equation for scalar field  and the energy density read
\begin{align}\label{sfd}
    \phi^{\prime 2}(y)=\frac{\text{csch}^2(t y) \left( 12^n k_1 n (2 n-1) (t\, \omega  \tanh (t y))^{2 n}+12 t^2 \omega ^2 \tanh ^2(t y)\right)}{4 \left(\frac{3 k_2}{2}+2\right) \omega },
\end{align}
and 
\begin{align}
    \rho(y)&=\text{sech}^{2 \omega }(t y)\bigg[\frac{\text{csch}^2(t y) \left(12^n k_1 n (2 n-1) (t\, \omega  \tanh (t y))^{2 n}+12 t^2 \omega ^2 \tanh ^2(t y)\right)}{16 \left(3 k_2+4\right) \omega }\nonumber\\&+\frac{12^n k_1 (2 n-1) (t\, \omega  \tanh (t y))^{2 n} \left(n\, \text{csch}^2(t y)-4 \omega \right)+12 t^2 \omega ^2 \left(\text{sech}^2(t y)-4 \omega  \tanh ^2(t y)\right)}{4 (5 k_2+4) \omega }\bigg].
\end{align}

Now, our task is to obtain the scalar field by solving Eq. (\ref{sfd}). For $n = 1$, the solution for the scalar field is given by
\begin{align}
    \phi(y)= \sqrt{\frac{24\, t\,\omega (k_1+1)}{3k_2+4}} \arctan \left[\tanh \left(\frac{t y}{2}\right)\right].
\end{align}
Such a solution represents a kink-like configuration illustrated in Fig. (\ref{fig1})(b). The matter field possesses a $Z_2$ symmetry and interpolates between the vacuum $\phi_0 = \pm \sqrt{\frac{24\, t\, \omega (k_1+1)}{3k_2+4}}$. It is possible to observe that the vacuum expectation value is influenced by parameters related to nonmetricity ($k_1$) and matter ($k_2$). On the other hand, the potential for this scalar field is obtained from Eq. (\ref{erer2}):
\begin{align}
  V(\phi) = \frac{3 (k_1+1)}{5 k_2+4} \bigg\{ 5 \, \text{sech}^2 \bigg[ \frac{2}{t} \, \tanh^{-1} \bigg( \tan \bigg( t \phi \sqrt{\frac{3k_2+4}{24t\omega(k_1+1)}} \bigg) \bigg) \bigg] - 4 \bigg\},
\end{align}with $\vert\phi\vert < \pi \sqrt{\frac{24\, t\, \omega (k_1+1)}{3k_2+4}}$. Additionally, the energy density is given by
\begin{align}
  \rho(y) = -\frac{6 (k_1+1) t^2 \omega \, \text{sech}^2(t y) \, \text{sech}^{2 \omega}(y) \left[ (3 k_2+4) \omega \, \cosh (2 t y) - k_2 (3 \omega +4) - 4 (\omega +1) \right]}{(3 k_2+4) (5 k_2+4)}.
\end{align}
As expected from the scalar field solution, the interaction potential exhibits a Higgs-like shape with two minima, which are influenced by the parameters $k_1$ and $k_2$. The energy density, depicted in Fig. (\ref{fig1})(d), also has a Higgs-like profile.

For $n=2$, we obtain the following analytical solution for the matter field
\begin{align}
   \phi(y)= \sqrt{\frac{3\, \omega\,\text{cosh}^2(ty)\text{sech}^2(ty) \left(72 k_1 t^2 \omega ^2 \tanh ^2(ty)+1\right)}{3 k_2+4}}\, \Gamma(ty),
\end{align}
where
\begin{align}
   \Gamma(ty)= \sqrt{2} \sinh (ty)+2 i \cosh (ty) (E(i t y, 72 k_1 t^2 \omega ^2+1)-F(i t y, 72 k_1 t^2 \omega ^2+1)),
\end{align}with $F(\varphi, m)$ and $E(\varphi, m)$ being the first and second kind elliptic integrals, respectively.

\begin{figure}[ht!]
\begin{center}
\begin{tabular}{ccc}
\includegraphics[height=5cm]{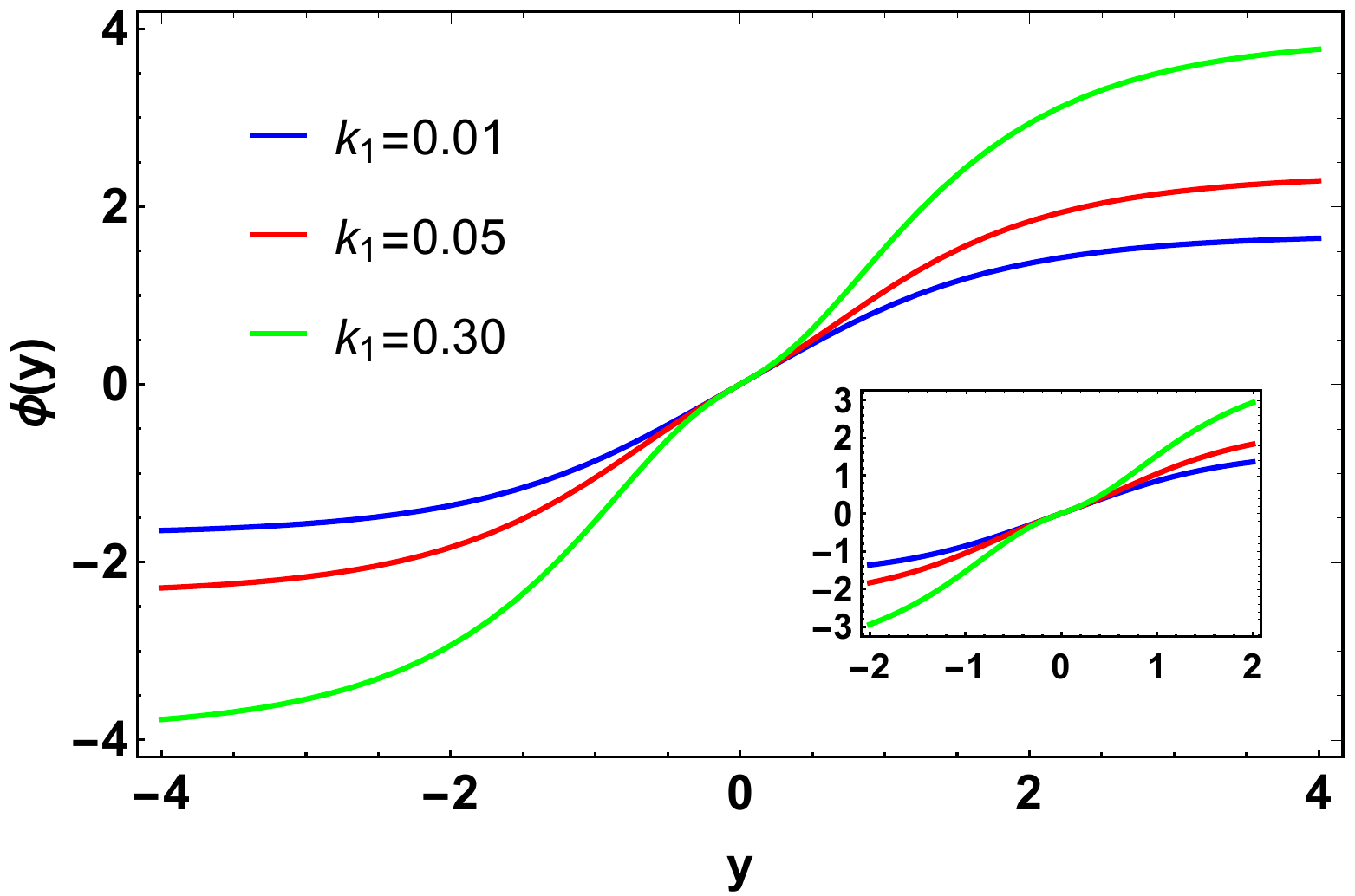} 
\includegraphics[height=5cm]{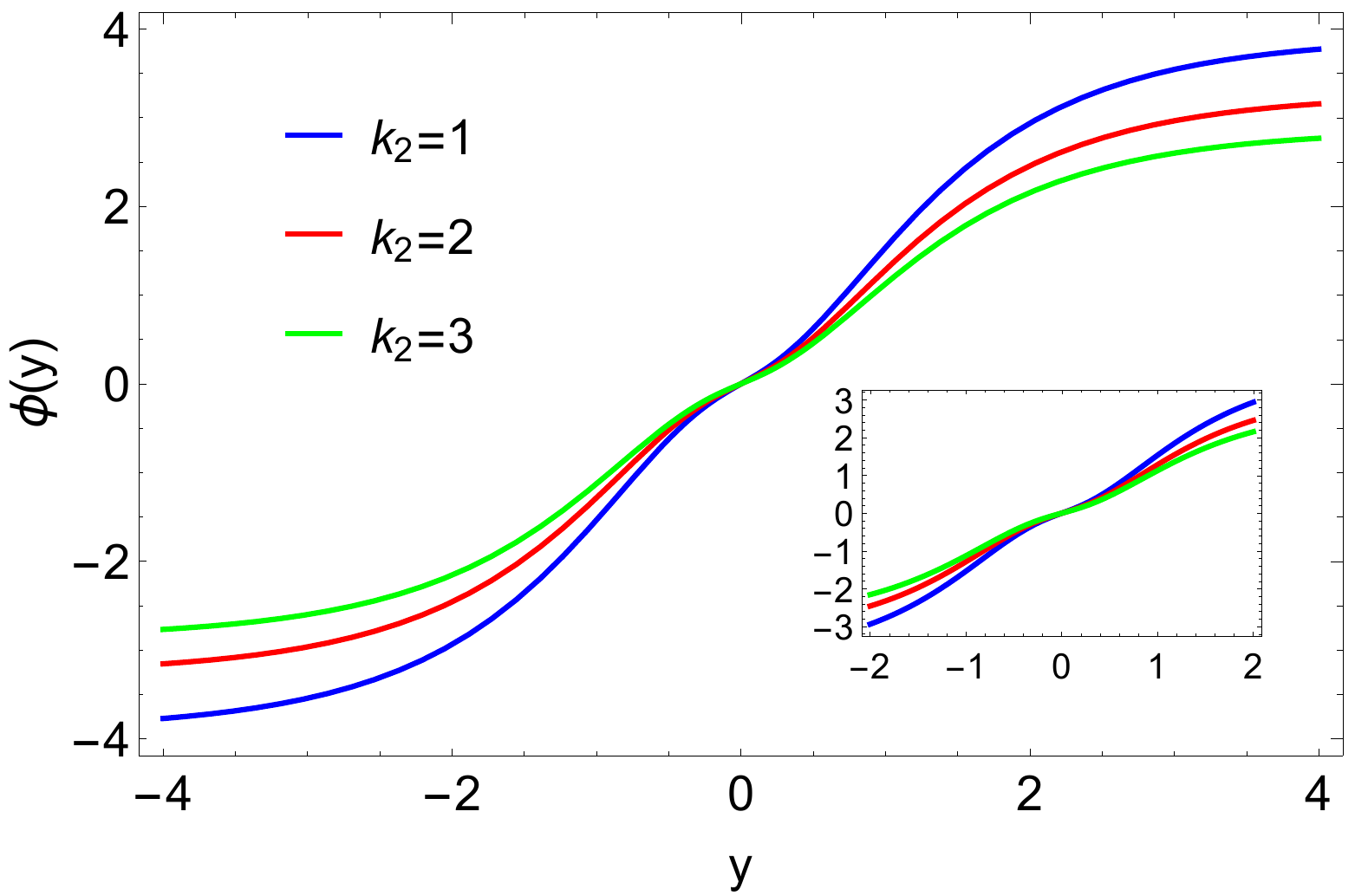}\\
(a)\hspace{6.7cm}(b)\\
\includegraphics[height=5cm]{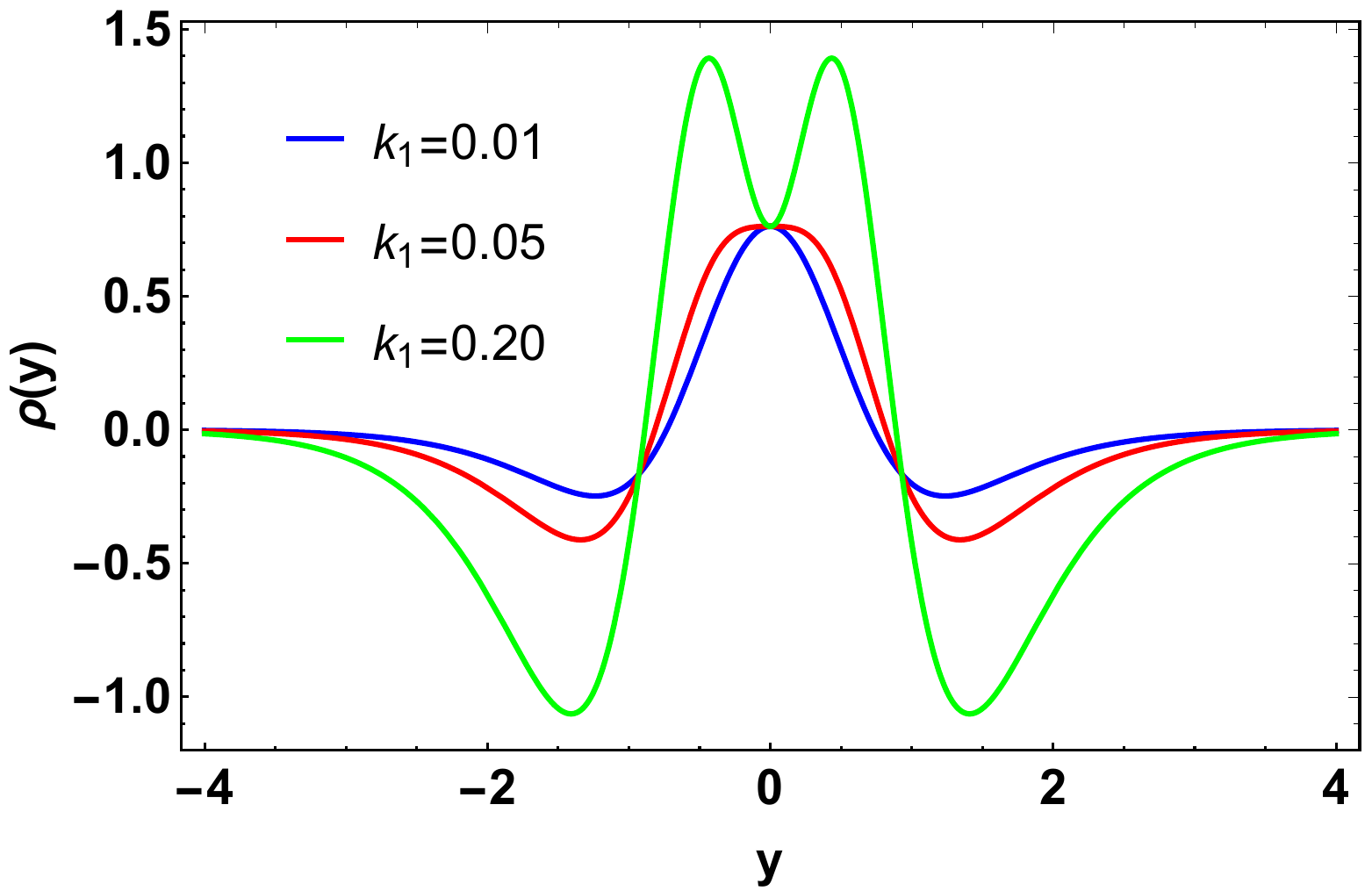}
\includegraphics[height=5cm]{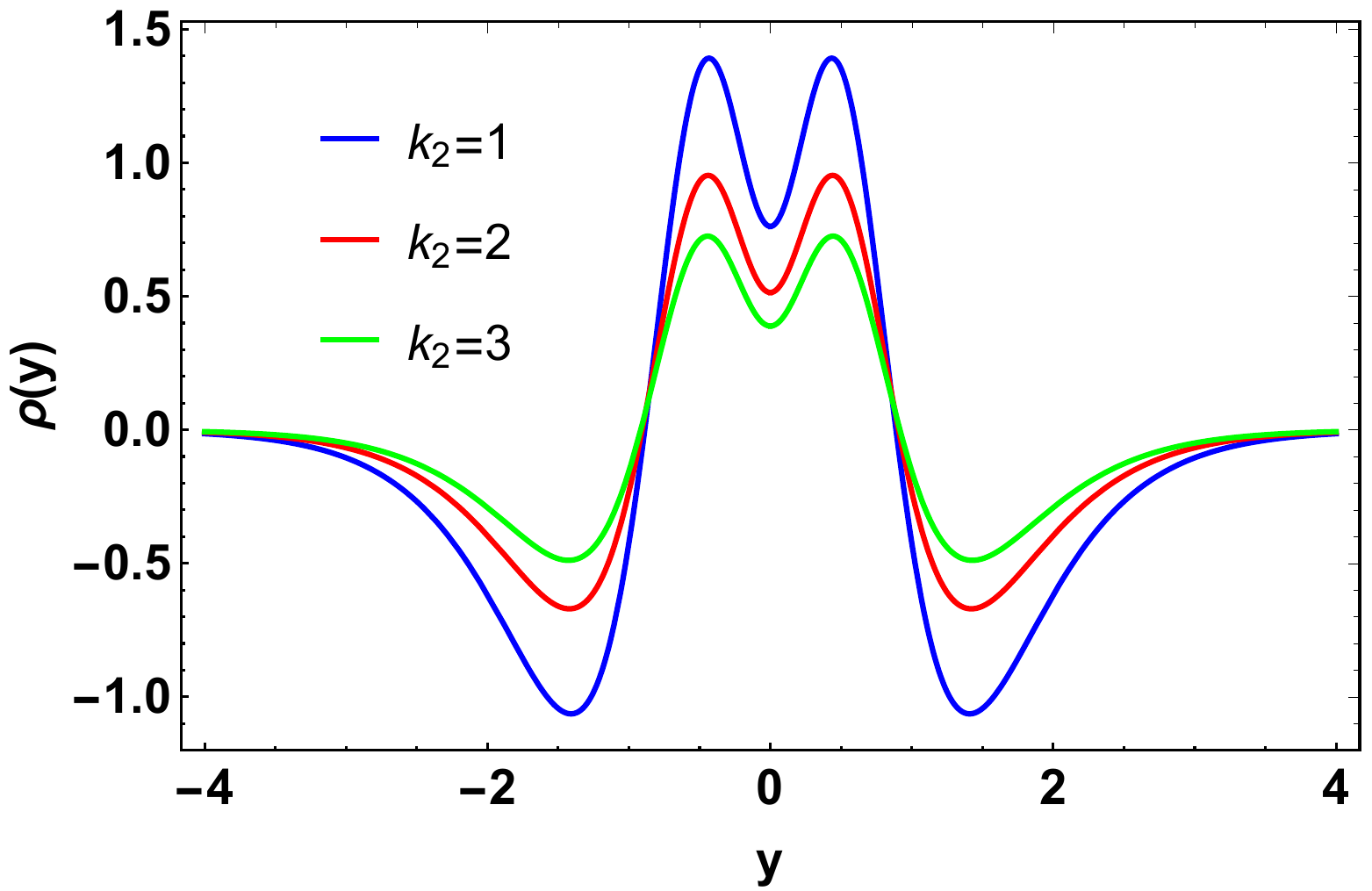}\\
(c)\hspace{6.7cm}(d) 
\end{tabular}
\end{center}
\vspace{-0.5cm}
\caption{For model 1 with $n=2$. (a) Scalar field with $k_2=1$. (b) Scalar field $k_1=0.2$. (c) Energy density $k_2=1$. (d) Energy density $k_1=0.2$.
\label{fig2}}
\end{figure}

The energy density now is written as
\begin{align}
    \rho(y)&=-\frac{12 t^2 \omega  \text{sech}^{2 \omega }(y) }{(3 \text{k2}+4) (5 k_2+4)}\bigg[(3 k_2+4) \omega  \tanh ^2(t y) \left(36 k_1 t^2 \omega ^2 \tanh ^2(t y)+1\right)\nonumber\\&-(k_2+1) \text{sech}^4(t y) \left(-72 k_1 t^2 \omega ^2+\left(72 k_1 t^2 \omega ^2+1\right) \cosh (2 t y)+1\right)\bigg].
\end{align}
We plot the shapes of $\phi(y)$ and $ \rho(y)$ in Fig. (\ref{fig2}) for some values of $k_1$ and $k_2$. Contrary to the case with $n=1$, the matter field solution for $n=2$ displays a behavior in which the kink-like profile is deformed into a two-kink structure, and the vacuum increases as $k_1$ increases. Such a behavior is also suggested by energy density which tends to split as $k_1$ increases. One can also note that the vacuum decreases as the parameter $k_2$ increases, similar to the previous case.

\begin{figure}[ht!]
\begin{center}
\begin{tabular}{ccc}
\includegraphics[height=5cm]{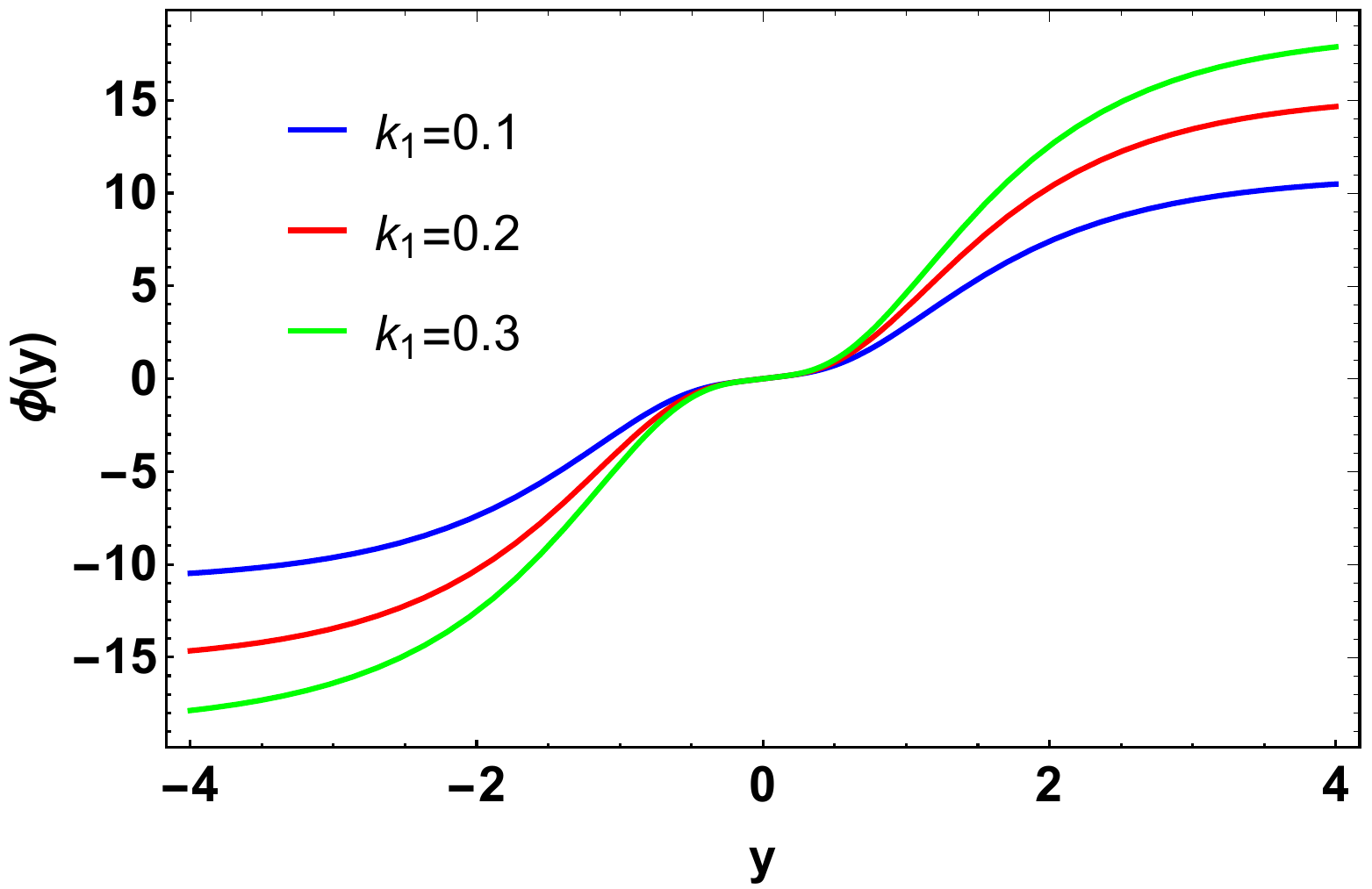} 
\includegraphics[height=5cm]{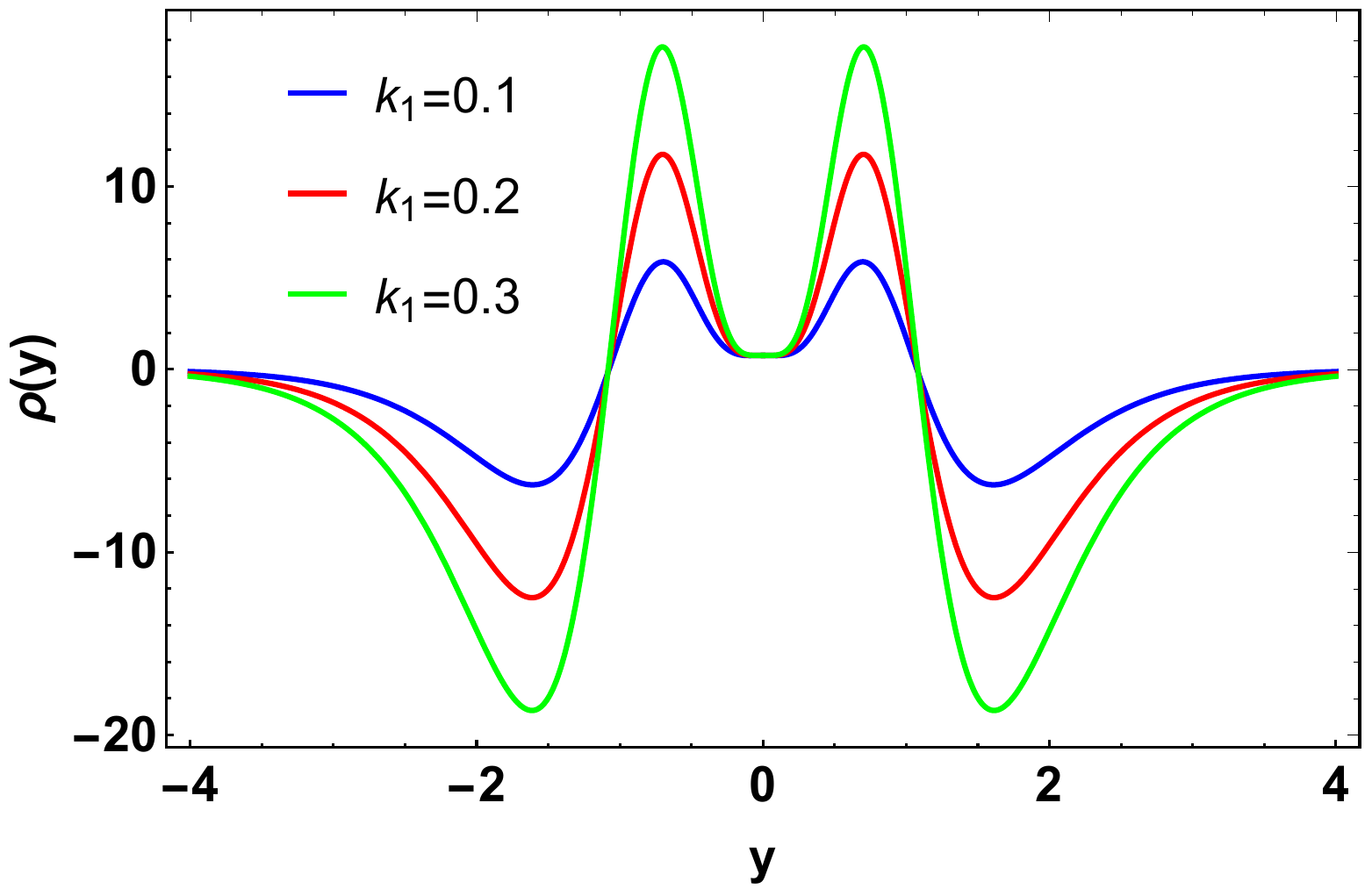}\\
(a)\hspace{6.7cm}(b)\\
\includegraphics[height=5cm]{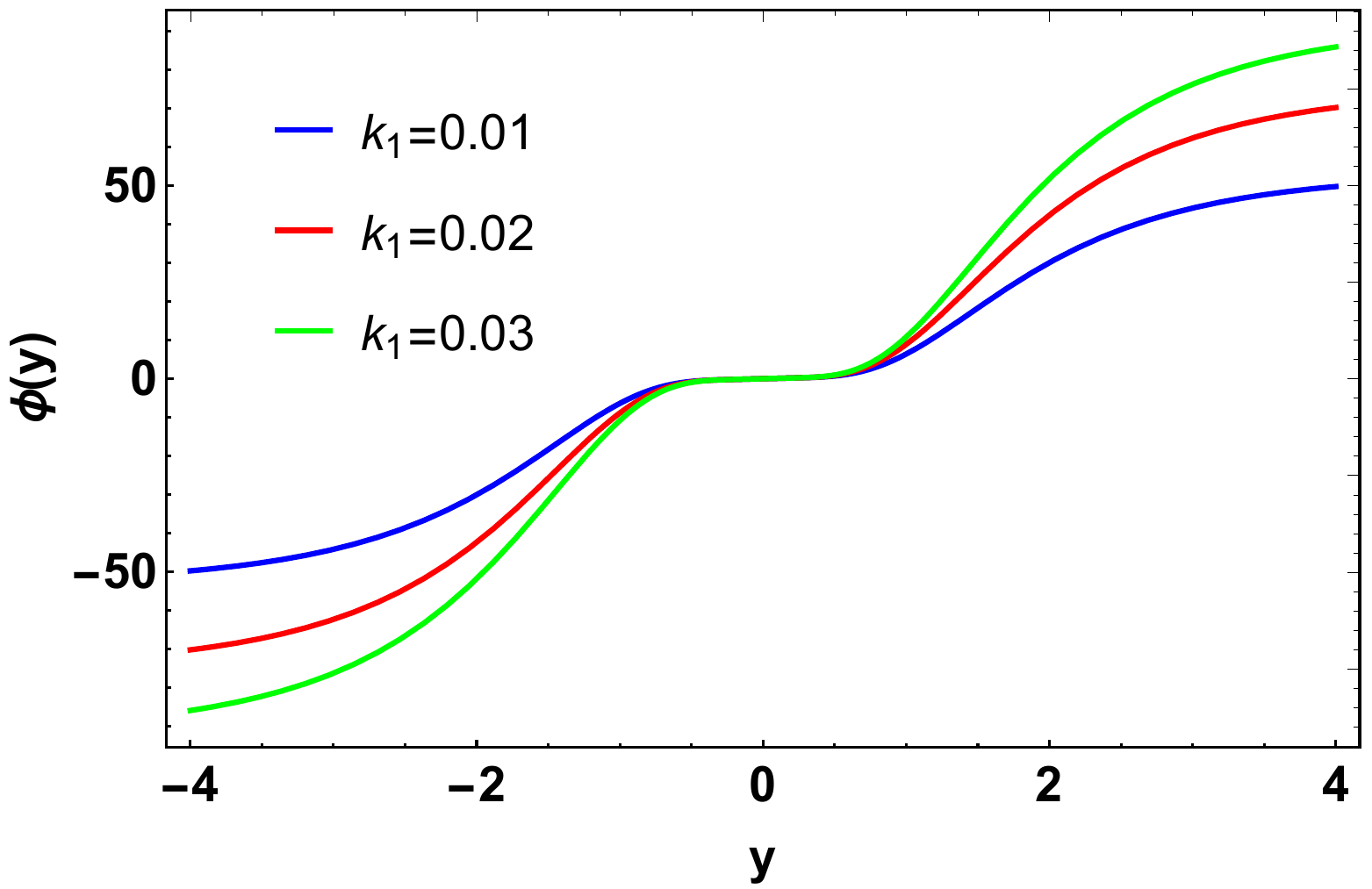}
\includegraphics[height=5cm]{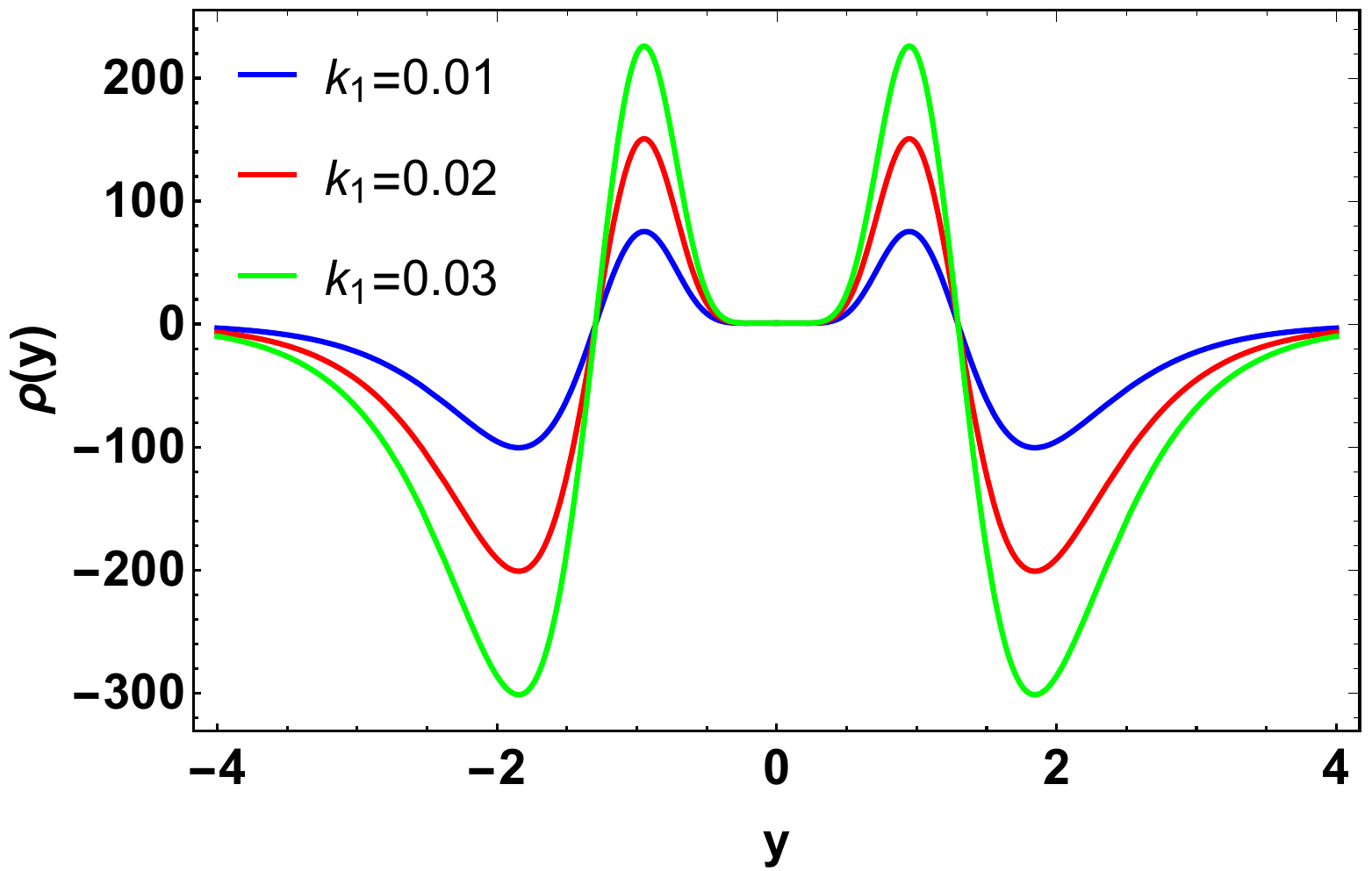}\\
(c)\hspace{6.7cm}(d) 
\end{tabular}
\end{center}
\vspace{-0.5cm}
\caption{ For model 1 with $n=3$. (a) Scalar field with $k_2=1$. (b) Scalar field $k_1=0.2$. (c) Energy density $k_2=1$. (d) Energy density $k_1=0.2$.
\label{fig3}}
\end{figure}

\begin{figure}[ht!]
\begin{center}
\begin{tabular}{ccc}
\includegraphics[height=5cm]{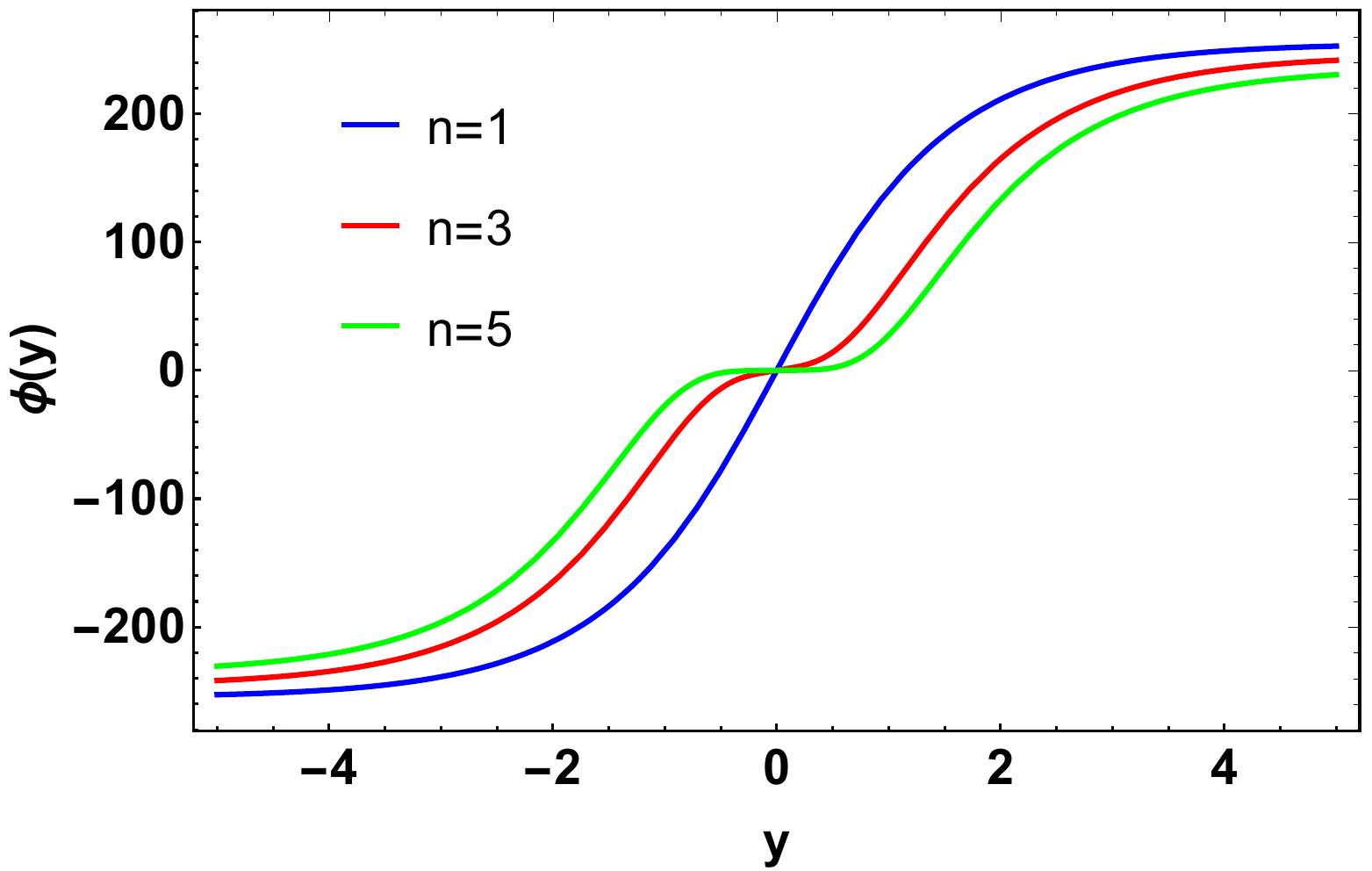} 
\includegraphics[height=5cm]{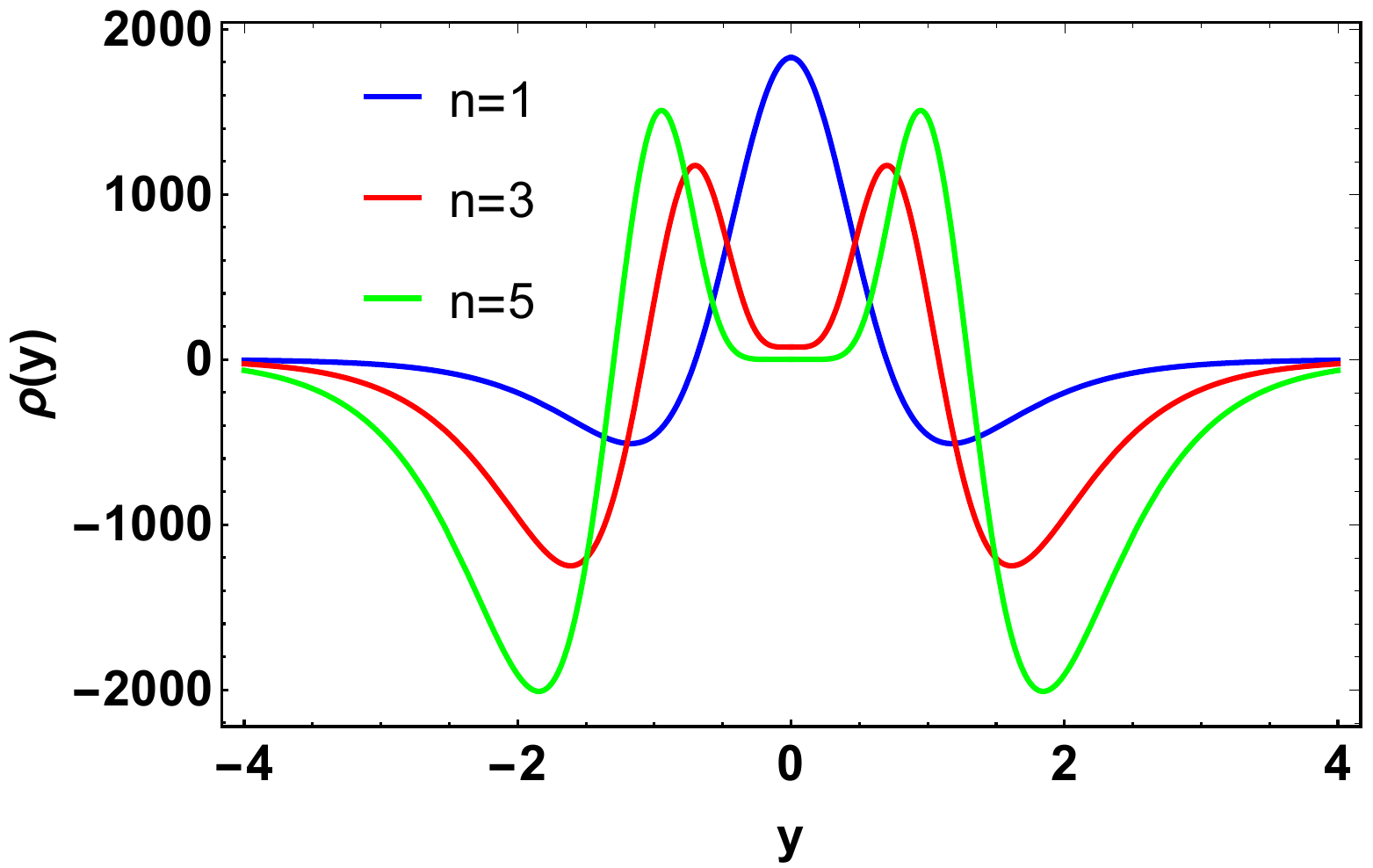}\\
(a)\hspace{6.7cm}(b)\\
\end{tabular}
\end{center}
\vspace{-0.5cm}
\caption{For model 1. (a) Scalar field with $k_1=0.2$ and $k_2=1$. (b) Energy density $k_1=0.2$ and $k_2=1$. 
\label{fig4}}
\end{figure}

For $n=3$ and $n=5$, the scalar field was obtained numerically. From Fig. (\ref{fig3}) and Fig. (\ref{fig4}), it is evident that the parameter $n$ plays a fundamental role in the deformation of the matter field, as well as in the splitting behavior observed in the energy density. The transition from a kink-like solution to a double-kink profile becomes more pronounced as $n$ and $k_1$ are modified. Furthermore, we can observe the influence of the parameter $k_2$ on the energy density, which decreases as $k_2$ increases.

\subsection{Model 2}

\begin{figure}[ht!]
\begin{center}
\begin{tabular}{ccc}
\includegraphics[height=5cm]{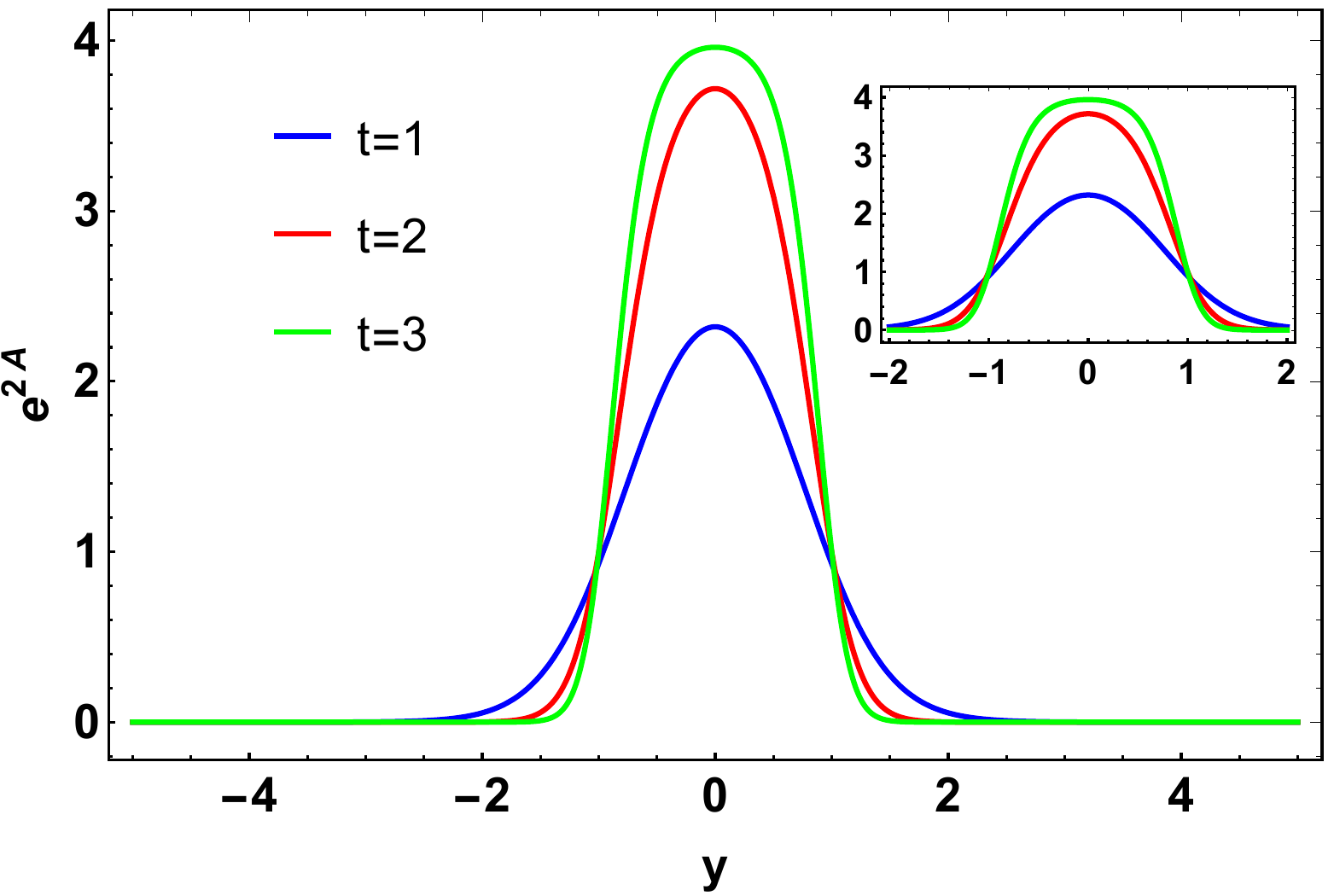} 
\includegraphics[height=5cm]{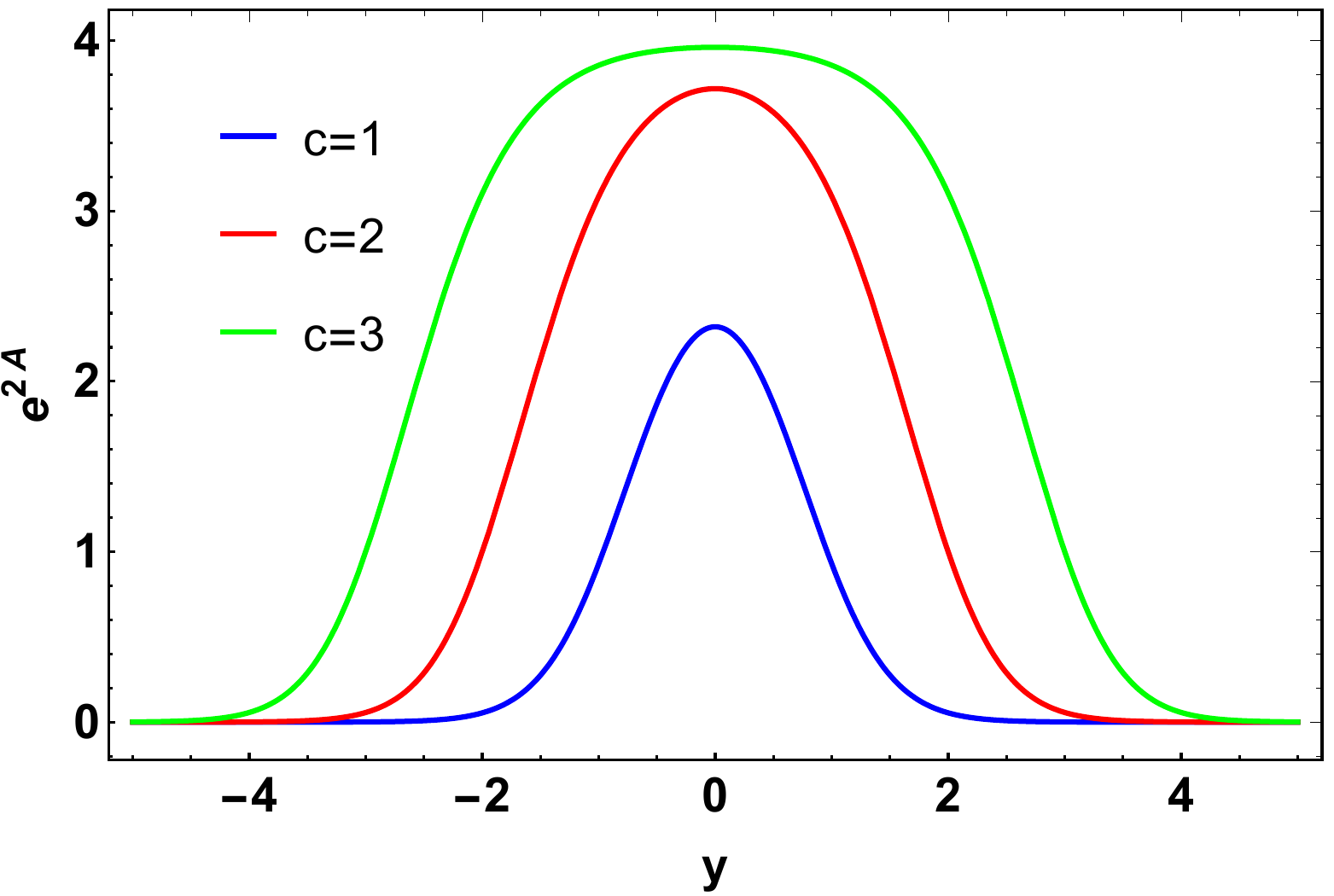}\\
(a)\hspace{6.7cm}(b)\\
\end{tabular}
\end{center}
\vspace{-0.5cm}
\caption{For model 2 with $n=2$. (a) Warp factor with $c=1$. (b) Warp factor $t=1$. 
\label{fig5}}
\end{figure}

Now, let us consider the second choice for the warp factor, namely
\begin{align}
    A(y)=\log{[\mathrm{tanh}[t(y+c)]-\mathrm{tanh}[t(y+c)]]}.
\end{align}  
The shape of this warp factor is illustrated in Fig. (\ref{fig5}). As observed, the warp factor exhibits a plateau near the brane core, which is controlled by the parameters $t$ and $c$. Specifically, the parameter $c$ regulates the width of the plateau, while the parameter $t$ determines its height.

\begin{figure}[ht!]
\begin{center}
\begin{tabular}{ccc}
\includegraphics[height=5cm]{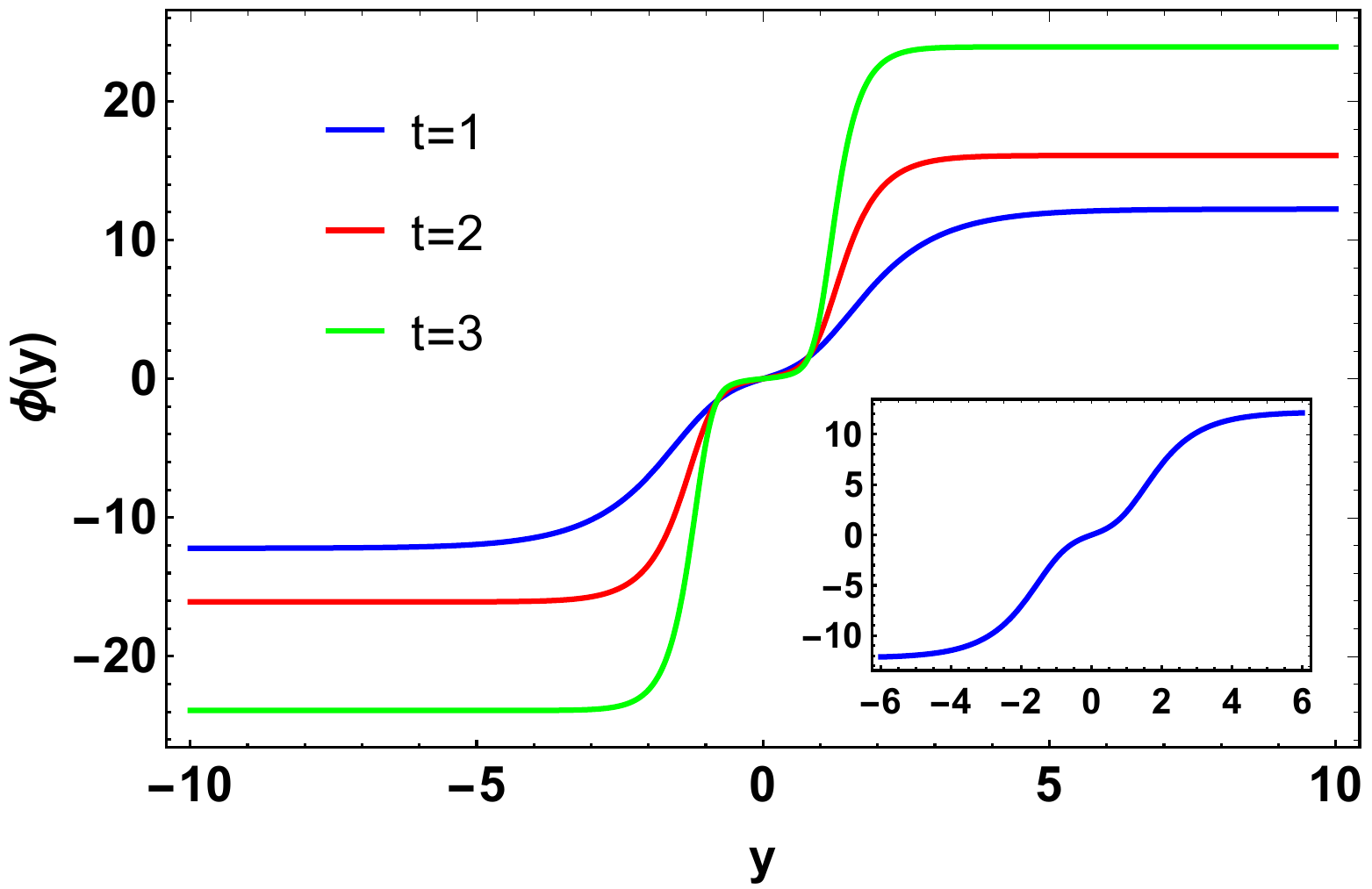} 
\includegraphics[height=5cm]{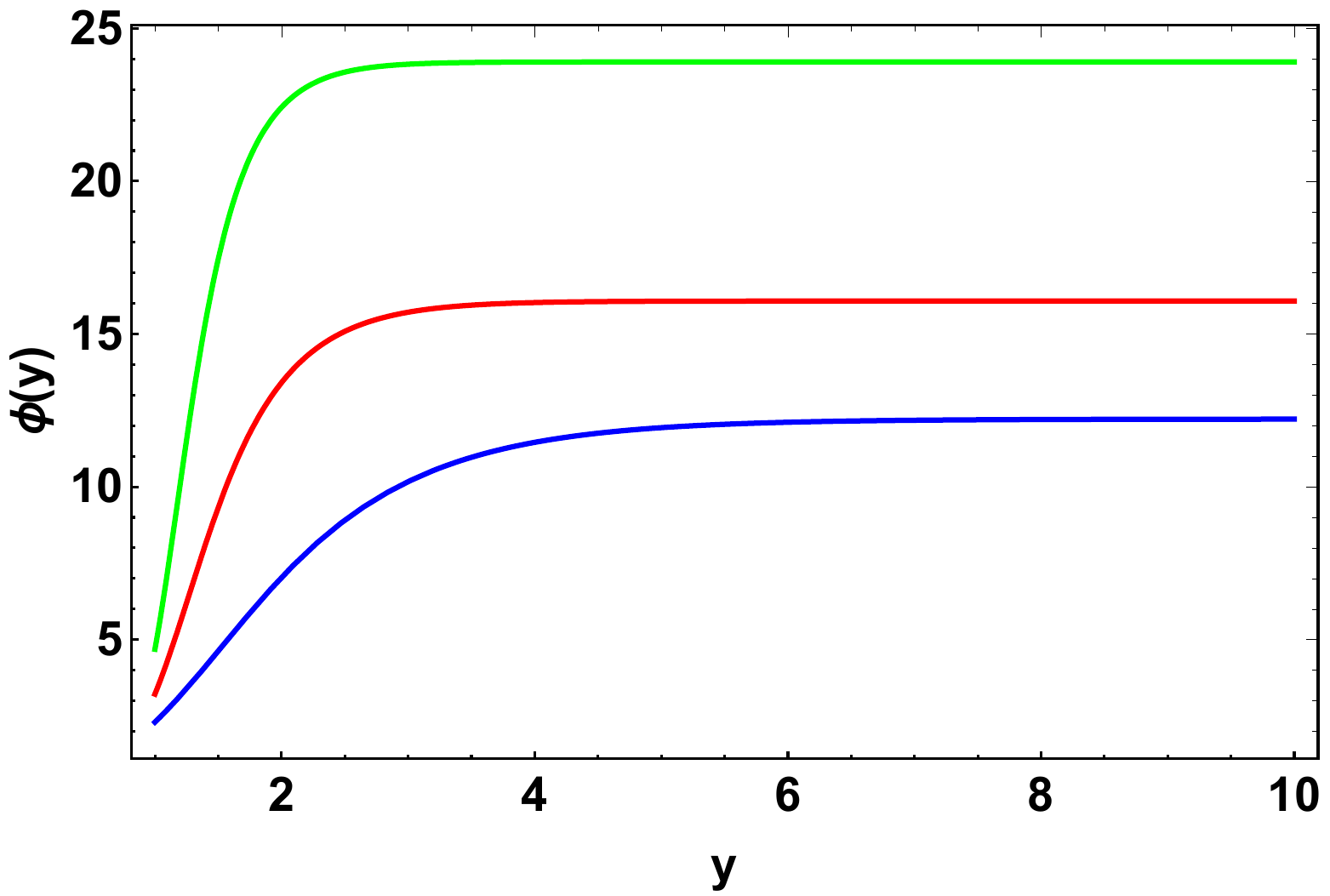}\\
(a)\hspace{6.7cm}(b)\\
\end{tabular}
\end{center}
\vspace{-0.5cm}
\caption{For model 2 with $n=2$. (a) Scalar field with $c=1$. (b) Scalar field with $c=1$. 
\label{fig6}}
\end{figure}

\begin{figure}[ht!]
\begin{center}
\begin{tabular}{ccc}
\includegraphics[height=5cm]{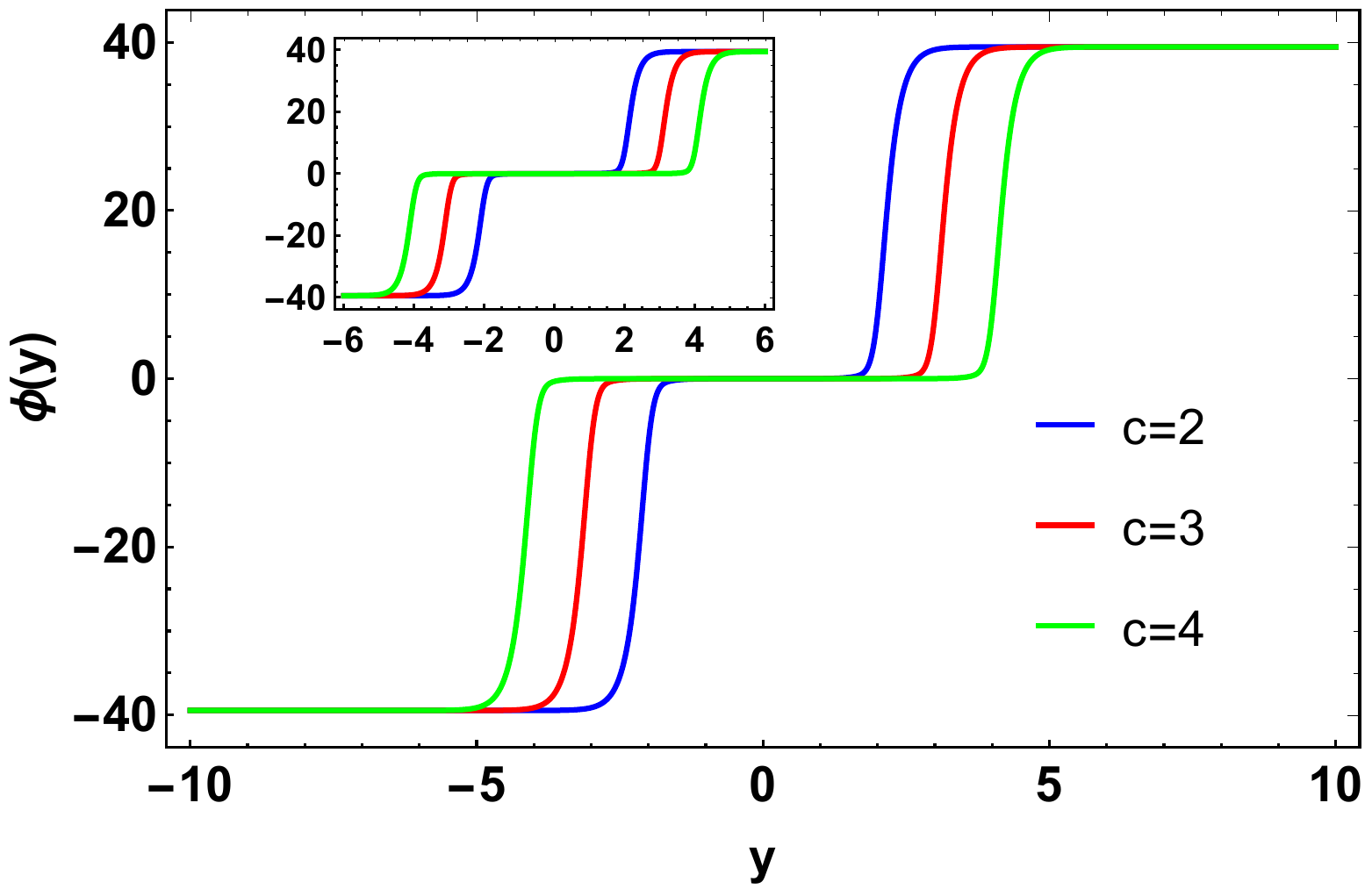} 
\includegraphics[height=5cm]{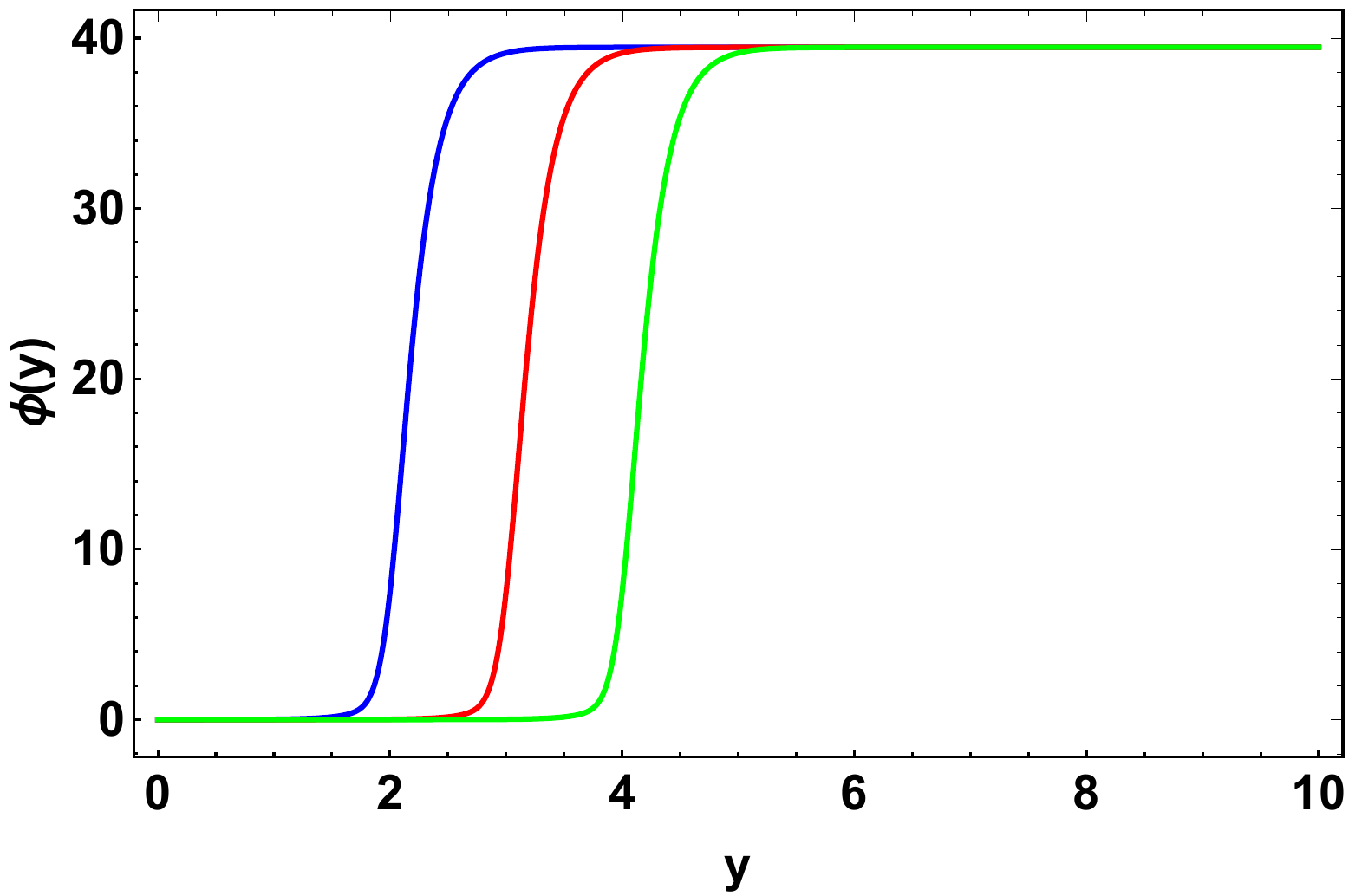}\\
(a)\hspace{6.7cm}(b)\\
\end{tabular}
\end{center}
\vspace{-0.5cm}
\caption{For model 2 with $n=2$. (a) Scalar field with $t=5$. (b) Scalar field with $t=5$. 
\label{fig7}}
\end{figure}

\begin{figure}[ht!]
\begin{center}
\begin{tabular}{ccc}
\includegraphics[height=5cm]{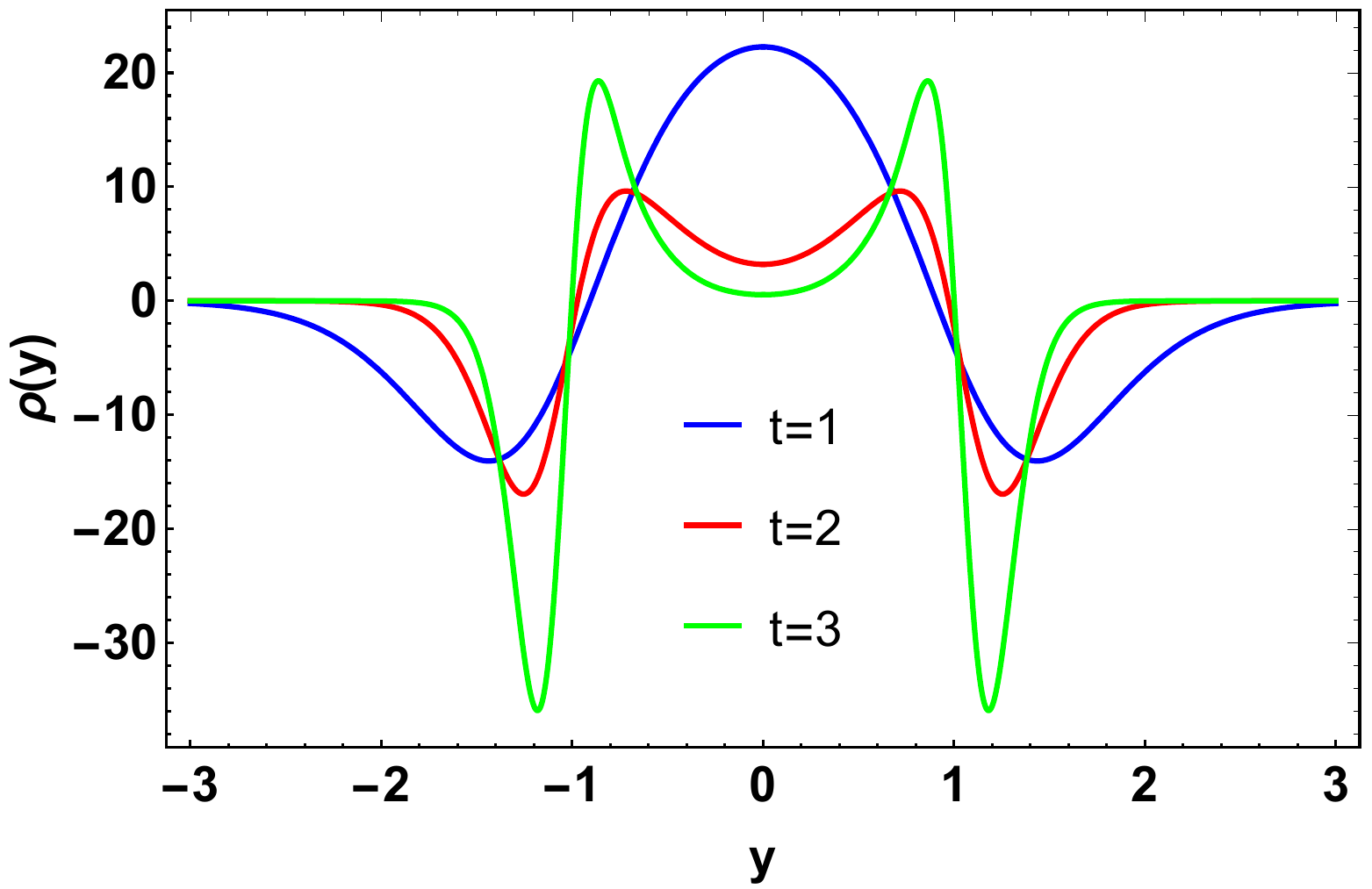} 
\includegraphics[height=5cm]{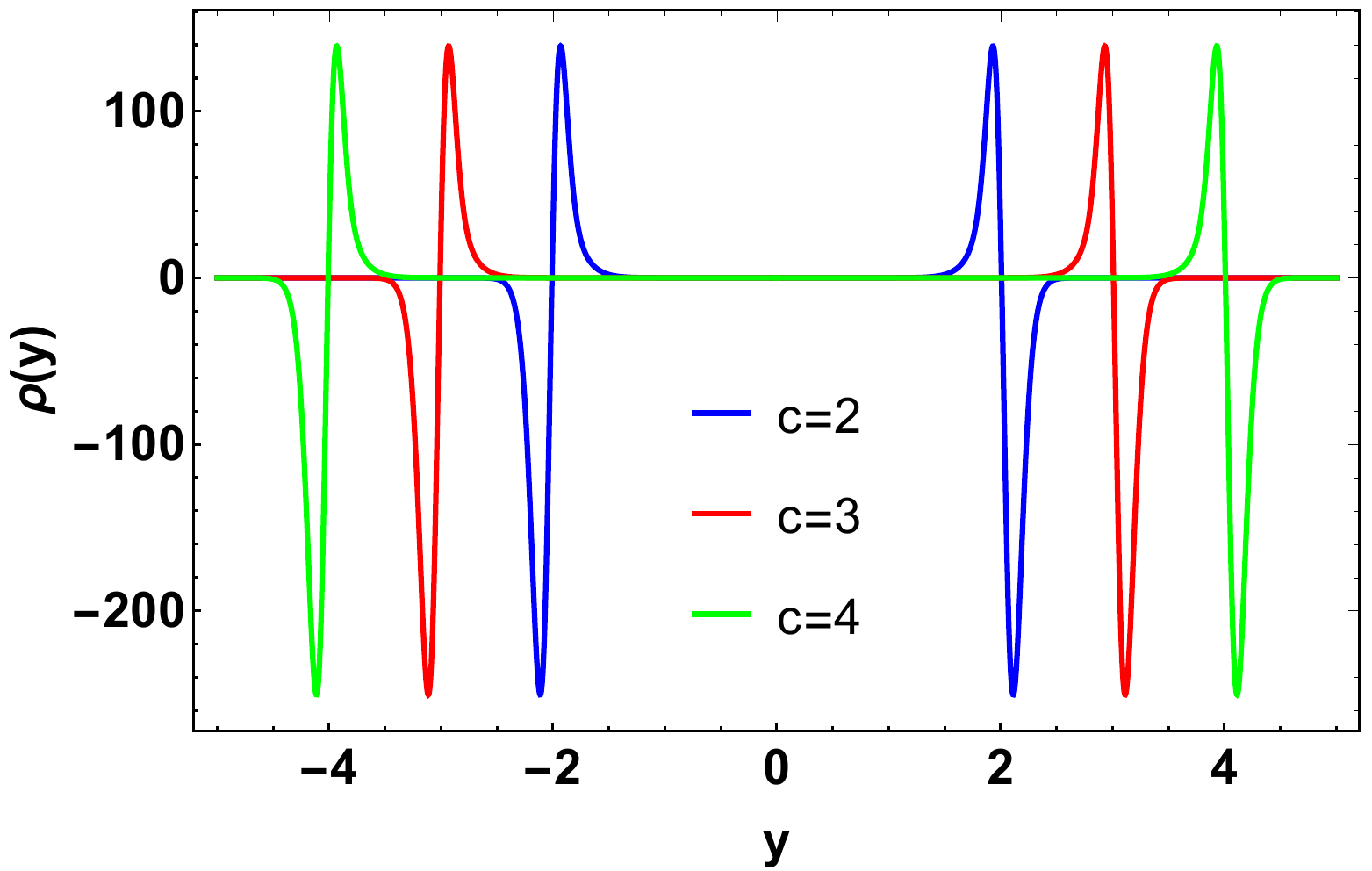}\\
(a)\hspace{6.7cm}(b)\\
\end{tabular}
\end{center}
\vspace{-0.5cm}
\caption{For model 2 with $n=2$. (a) Energy density with $c=1$. (b) Energy density with $t=5$.
\label{fig8}}
\end{figure}

After defining the warp factor, the next step is to determine the matter field. Given the complexity of the equation governing the matter field, we solve it numerically. Here, we do not explicitly present this equation or the corresponding expression for the energy density. Instead, their behavior is illustrated in Figs. (\ref{fig6}-\ref{fig8}). It is observed that the scalar field exhibits a compact double-kink structure as the parameter $t$ increases. Meanwhile, the parameter $c$ regulates the width of the compact region near the origin. Similarly, the energy density profile also displays a compact structure.

\section{Tensor perturbation and stability}\label{s4}

Once we have discussed the matter sector, let us turn our attention to the stability of the braneworld by considering small metric perturbations in the form $\eta_{\mu\nu}\rightarrow\eta_{\mu\nu}+h_{\mu\nu}(x^\mu,y)$ and small scalar field perturbations as $\phi \rightarrow \phi(y) + \epsilon(x^\mu,y)$, where $x^\mu$ denotes the position four-vector. It is worth highlighting that $h_{\mu\nu}$ represents the tensor propagation, which can be identified as the graviton. To simplify our analysis, we impose the gauge conditions $\partial^\mu h_{\mu\nu} = 0$ and $h = \eta^{\mu\nu} h_{\mu\nu}=0$, known as the transverse and traceless conditions, respectively. The metric incorporating tensor perturbations is given by
\begin{align}\label{metric2}
ds^2= \e^{2A}[\eta_{\mu\nu}+h_{\mu\nu}(x^\mu,y)]dx^{\mu}dx^{\nu}+dy^2.   
\end{align}

By considering the perturbed metric (\ref{metric2}), we obtain the nonvanishing nonmetricity tensor, explicitly \cite{Fu:2021rgu}
\begin{align}
&Q^{(1)\rho}\ _{\mu\nu}=\partial^\rho h_{\mu\nu},\\
&Q^{(1)4}\ _{\mu\nu}=e^{2A}(2A^{\prime}h_{\mu\nu}+h^{\prime}_{\mu\nu}),
\end{align}
while the nonvanishing perturbed nonmetricity conjugate is given by
\begin{align}
&P^{(1)\rho}\ _{\mu\nu}=-\frac{1}{4}\Big[\partial^\rho h_{\mu\nu}-(\partial_\mu h^\rho_\nu+\partial_\nu h^\rho_\mu)\Big],\\
&P^{(1) 4}\ _{\mu\nu}=\frac{1}{4}e^{2A}(6A^{\prime}h_{\mu\nu}-h^{\prime}_{\mu\nu}),\\
&P^{(1)\rho}\ _{4\nu}=P^{(1)\rho}\ _{\nu 4}=\frac{1}{4}h^{\prime \rho}_\nu.
\end{align}
Additionally, by considering the scalar pertubation, we write the perturbation of the energy-momentum tensor as 
\begin{align}
\mathcal{T}^{(1)}\ _{\mu\nu}= -\e^{2A}\Big(\frac{1}{2}\overline{\phi}^{\prime\, 2}h_{\mu\nu}+\overline{\phi}^{\prime}\epsilon^{\prime}\eta_{\mu\nu}+V h_{\mu\nu}+V_{\phi}\epsilon\eta_{\mu\nu}\big).  
\end{align}

Having established these expressions, let us obtain the equation of motion for $h_{\mu\nu}$ by employing the perturbed form of the gravitational equation (\ref{graeq}), leading to
\begin{align}\label{779}
h_{\mu\nu}^{\prime\prime}+\left(4A^{\prime}+\frac{f^{\prime}_Q}{f_Q}\right)h_{\mu\nu}^{\prime}=\e^{-2A}\square h_{\mu\nu}. 
\end{align}

To proceed further, let us employ the Kaluza Klein (KK) decomposition for graviton as follows:
\begin{align}\label{777}
h_{\mu\nu}(x^\lambda,y)=\sum_n \widehat{h}_{\mu\nu,\,(n)}(x^\lambda)\chi_n(y), 
\end{align}
where the 4D tensor satisfies the equation of motion $\square \widehat{h}_{\mu\nu,\,(n)}=m^2\widehat{h}_{\mu\nu,\,(n)}$. By substituting (\ref{777}) into (\ref{779}), we obtain the following equation for the scalar $\chi(y)$
\begin{align}\label{111}
\chi^{\prime\prime}+\left(4A^{\prime}+\frac{f^{\prime}_Q}{f_Q}\right)\chi^{\prime}=-m^2\e^{-2A}\chi.
\end{align}
At this point, it is appropriate to introduce the conformal coordinate $dz=\e^{-A}dy$, so that the Eq. (\ref{111}) is rewritten as follows:
\begin{align}\label{113}
\ddot{\chi}(z)+2H\dot{\chi}(z)=-m^2\chi(z),
\end{align}
where
\begin{align}
H=\frac{1}{2}\Big(3\dot{A}+\frac{\dot{f}_Q}{f_Q}\Big).
\end{align}

Henceforth, the dot (\ $\dot{ }$\ ) denotes differentiation with respect to the conformal coordinate $z$. To move forward, we rewrite Eq. (\ref{113}) in a Schrödinger-like form. For this purpose, we apply the transformation $\chi(z) = e^{K(z)}\psi(z)$, where $K = -\int H dz$. After some manipulation, we obtain the following Schrödinger-like equation:
\begin{align}\label{114}
-\ddot{\psi}+V\psi=m^2\psi,
\end{align}
where we have defined the effective potential
\begin{align}
V=\dot{H}+H^2.
\end{align}
The equation (\ref{114}) is a Schrödinger-like equation that arises from supersymmetric quantum mechanics. Expressing Eq. (\ref{114}) in this form is important since it guarantees us the stability of the graviton spectrum. Moreover, it is possible to rewrite Eq. (\ref{114}) in a factorized form:
\begin{align}
\bigg(\partial_z+\frac{3}{2}\dot{A}+\frac{1}{2}\frac{\dot{f}_Q}{f_Q}\bigg)\bigg(-\partial_z+\frac{3}{2}\dot{A}+\frac{1}{2}\frac{\dot{f}_Q}{f_Q}\bigg)\psi=m^2\psi.  \end{align}

It is important to note that there is no tensor tachyon mode with $m^2 < 0$, confirming that the brane is indeed stable against tensor perturbations. Another advantage of this approach is that it enables us to express the massless mode (zero-mode) as  
\begin{align}\label{eqcampo}
\psi_0(z)=N_0\e^{\frac{3}{2}A+\frac{1}{2}\int\frac{\dot{f}_Q}{f_Q}dz},
\end{align}
where $N_0$ is a normalization constant, which can be fixed through the condition
\begin{align}
\int_\infty^\infty dz\, \psi_0(z)=\int_\infty^\infty dy\, \e^{-A(y)}\, \psi_0(y)=1.    \end{align}

The next step is to study the gravity localization for the two warp factors analyzed in the previous section. 

\subsection{Model 1}
For the warp factor (\ref{wf1}), we can use the conformal coordinate $dz=\e^{-A}dy$ and write
\begin{align}
y=\log \left(\text{sech}\left(\sinh ^{-1}(z)\right)\right),
\end{align}
so that the warp factor as a function of the conformal coordinate reads
\begin{align}
 A(z)=-\frac{1}{2} \log \left(z^2+1\right).   
\end{align}
With this, the simplest case we can analyze is for $n = 1$, where the zero-mode and the effective potential are given explicitly:
\begin{align}
\psi_0(z)=N_0 \e^{\frac{3}{2}A(z)}=\frac{N_0}{\left(z^2+1\right)^{3/4}},    
\end{align}
and
\begin{align}
V(z)=\frac{3 \left(5 z^2-2\right)}{4 \left(z^2+1\right)^2}.    
\end{align}

\begin{figure}[ht!]
\begin{center}
\begin{tabular}{ccc}
\includegraphics[height=5.5cm]{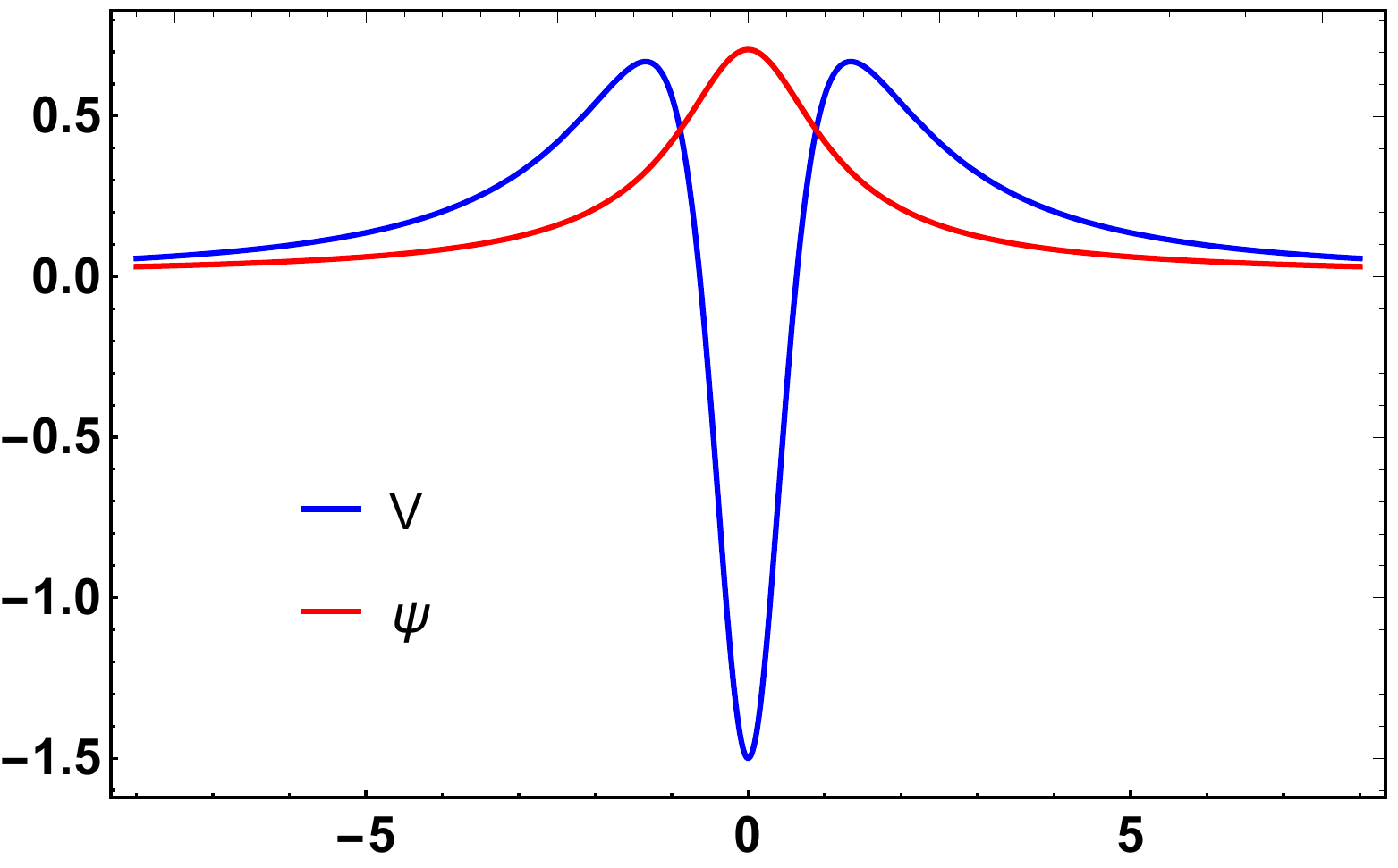} 
\end{tabular}
\end{center}
\vspace{-0.5cm}
\caption{ For model 1 with $n=1$. Effective potential (blue line) and zero-mode (red line)
\label{fig9}}
\end{figure}

In this case, we have $N_0=\sqrt{\frac{1}{2}}$. We expose the behaviors of both the effective potential and the zero mode in Fig. (\ref{fig9}), and as we can see, the parameter does not affect both the zero mode and the effective potential. The zero mode is well localized with a Gaussian profile, while the effective potential exhibits a volcano-like shape. 

On the other hand, for $n=2$ we find 
\begin{align}
\psi_0(z)= \frac{1}{\left(z^2+1\right)^{5/4}}\sqrt{\frac{1+(24 k_1+1)z^2}{16 k_1+2}},
\end{align}
and
\begin{align}
V(z)=\frac{3 \left[8 \left(1-720 k_1^2+6 k_1\right) z^4+5 (24 k_1+1)^2 z^6+(1-256 k_1) z^2+32 k_1-2\right]}{4 \left(z^2+1\right)^2 \left[(24 k_1+1) z^2+1\right]^2}.   
\end{align}

\begin{figure}[ht!]
\begin{center}
\begin{tabular}{ccc}
\includegraphics[height=5cm]{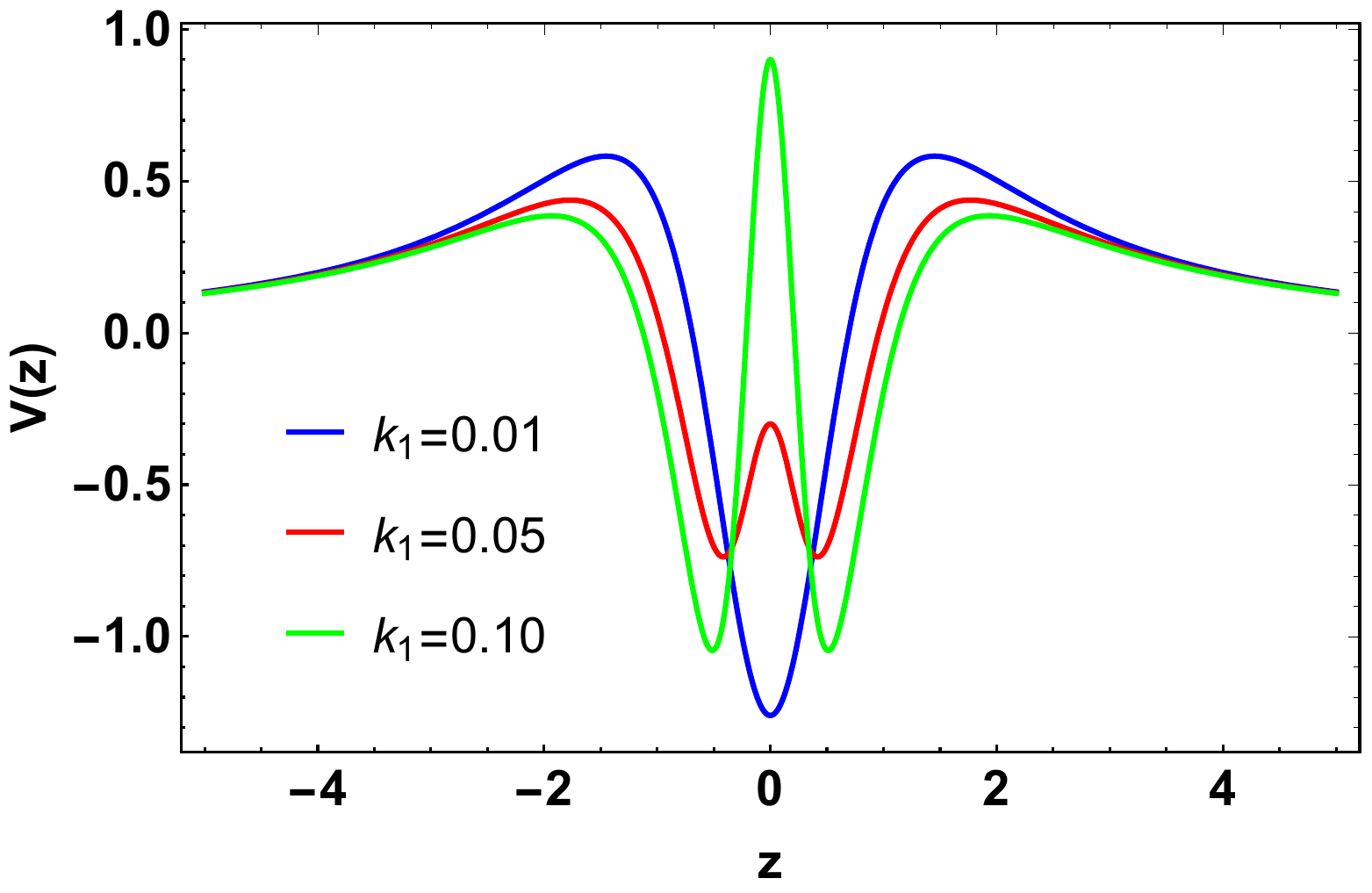} 
\includegraphics[height=5cm]{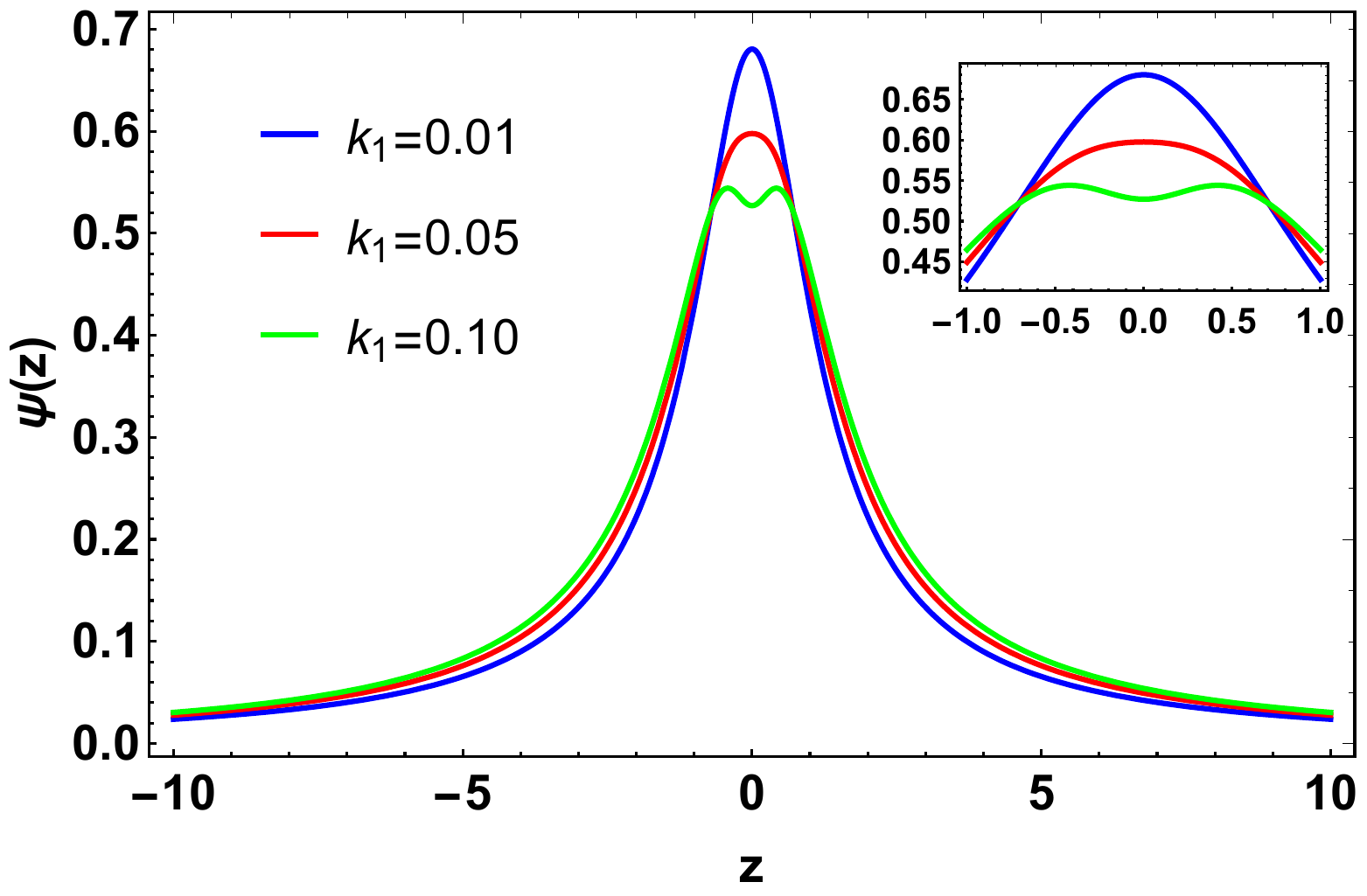}\\
(a)\hspace{6.7cm}(b)\\
\end{tabular}
\end{center}
\vspace{-0.5cm}
\caption{ For model 1 with $n=2$. (a) Effective potential. (b) Zero-mode
\label{fig10}}
\end{figure}

Unlike in the previous case, we see the influence of the nonmetricity parameter on both the zero-mode and the effective potential. The Fig. (\ref{fig10}) shows their behavior. For instance, in Fig. (\ref{fig10})(a), we can observe the formation of two symmetric wells as a consequence of brane splitting, while in Fig.(\ref{fig10})(b), the zero-mode tends to present two peaks (split) as the parameter $k_1$ increases.

The last case to be considered in the first model is $n=3$, where we obtain
\begin{align}
\psi_0(z)= \frac{1}{\left(z^2+1\right)^{7/4}}\sqrt{\frac{5 \left[1+2z^2+(432 k_1+1)z^4 \right]}{864 k_1+10}},
\end{align}
and
\begin{align}
V(z)=\frac{7 \left(z^2-10\right)}{4 \left(z^2+1\right)^2}+\frac{2 \left[(432 k_1+1)^2 z^6+10 (432 k_1+1) z^4+(1296 k_1+17) z^2+8\right]}{\left[(432 k_1+1) z^4+2 z^2+1\right]^2}.
\end{align}

\begin{figure}[ht!]
\begin{center}
\begin{tabular}{ccc}
\includegraphics[height=5cm]{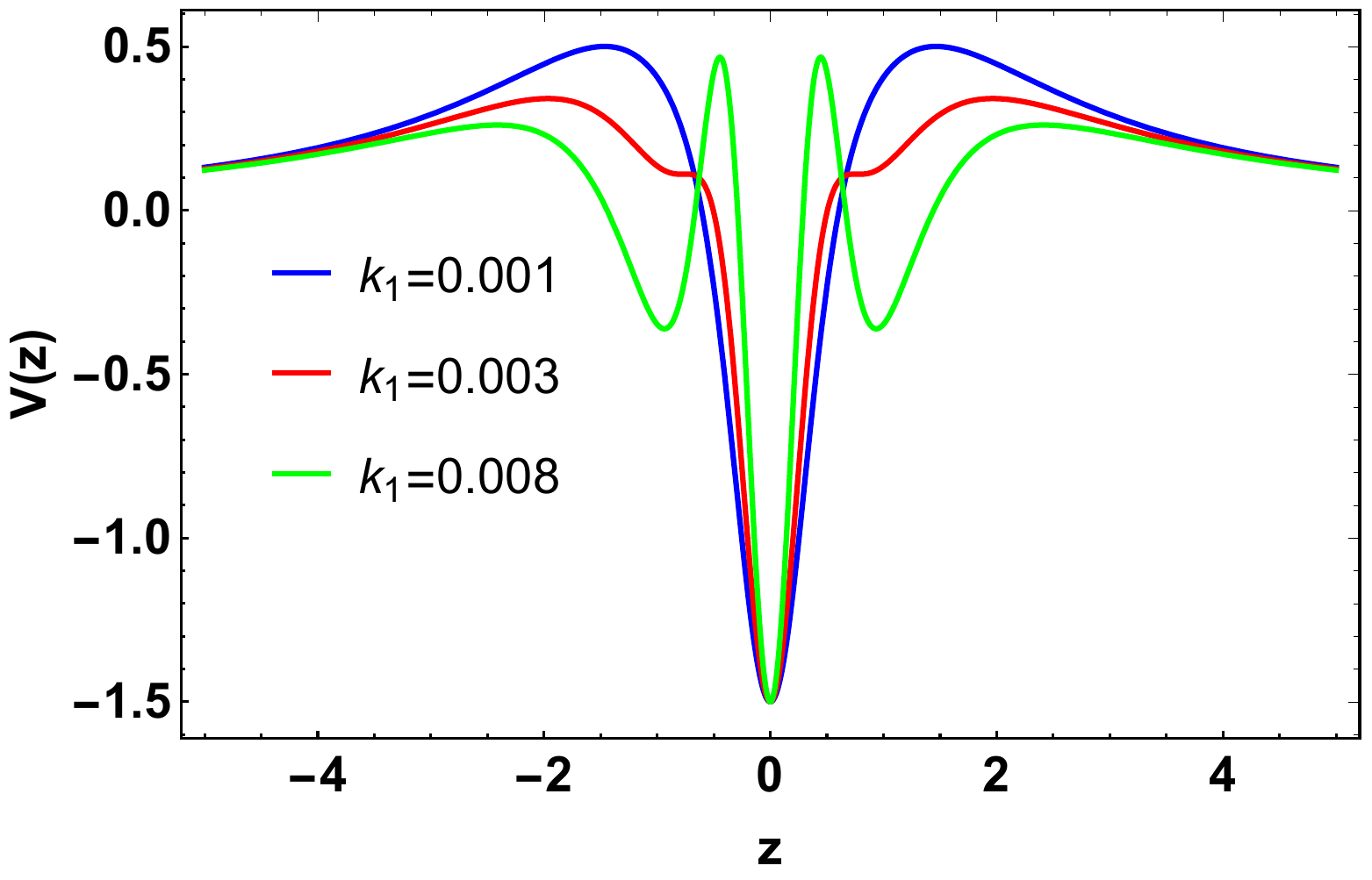} 
\includegraphics[height=5cm]{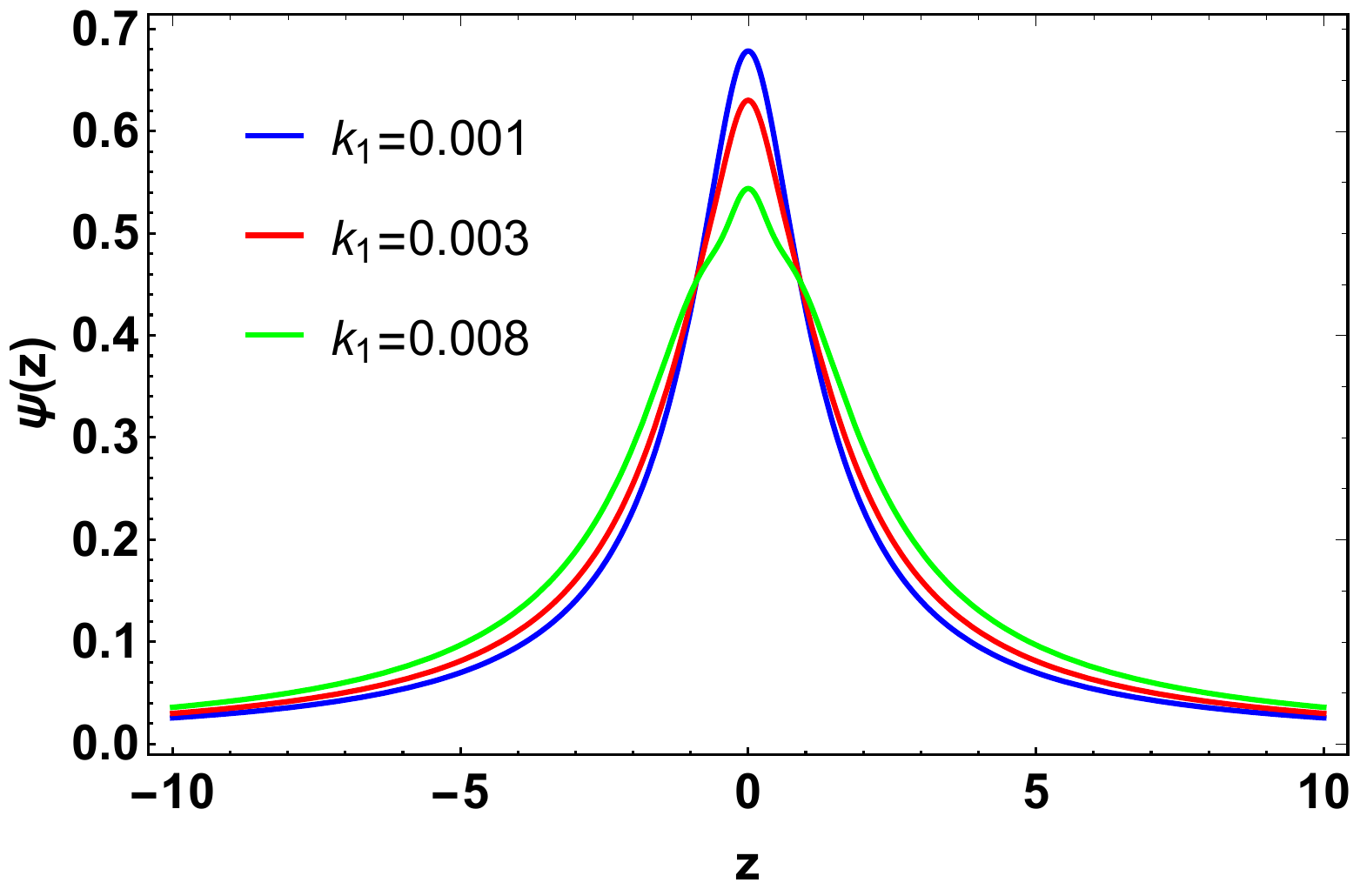}\\
(a)\hspace{6.7cm}(b)\\
\end{tabular}
\end{center}
\vspace{-0.5cm}
\caption{ For model 1 with $n=3$. (a) Effective potential. (b) Zero-mode
\label{fig11}}
\end{figure}

We again observe the influence of the nonmetricity parameter on both the zero-mode and the effective potential, as illustrated by Fig. (\ref{fig11}). In this case, we notice the emergence of a potential with three wells as $k_1$ increases: a deeper well located right at the origin and two symmetrical wells nearby.

\subsection{Model 2}

\begin{figure}[ht!]
\begin{center}
\begin{tabular}{ccc}
\includegraphics[height=5cm]{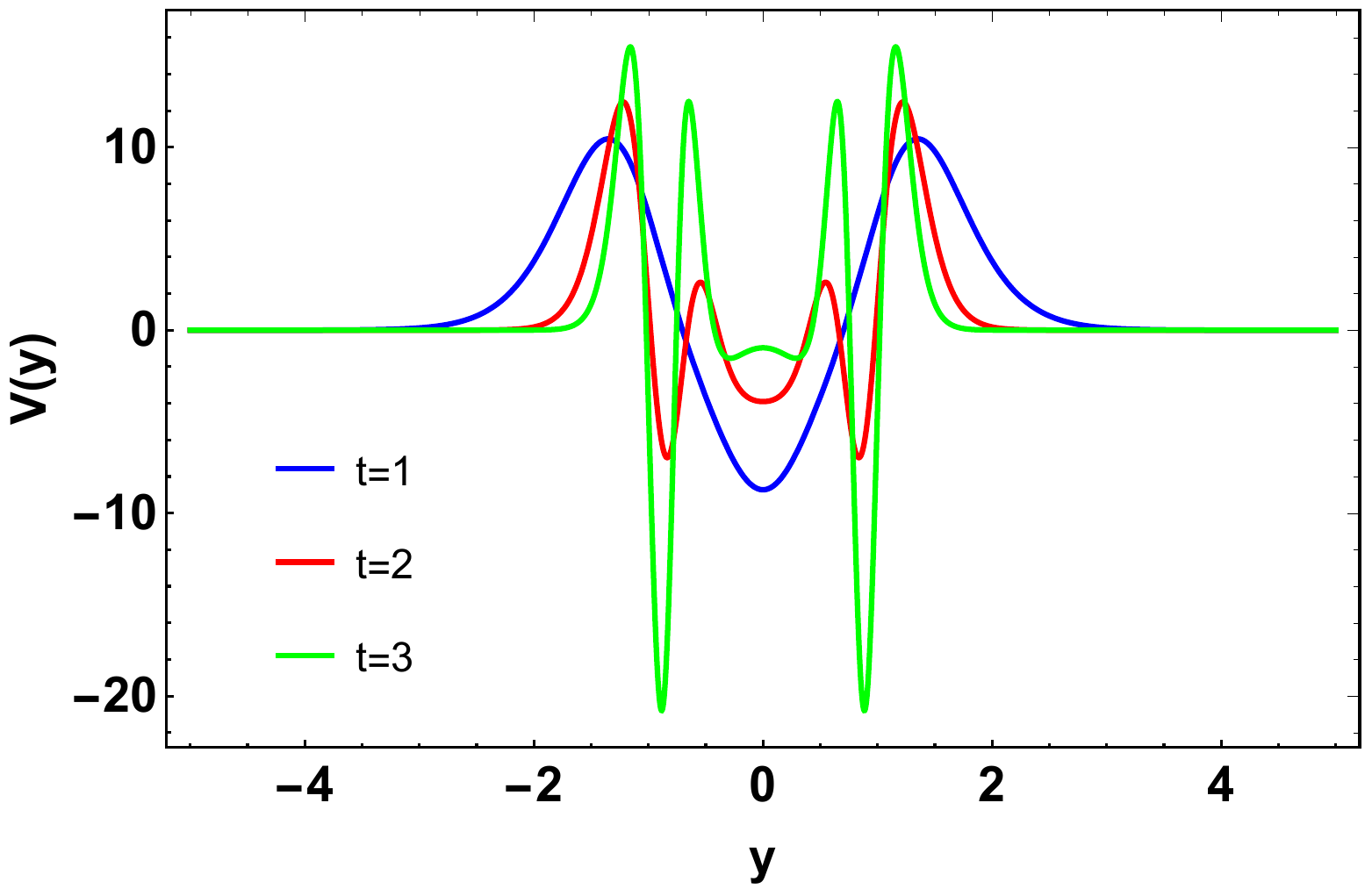} 
\includegraphics[height=5cm]{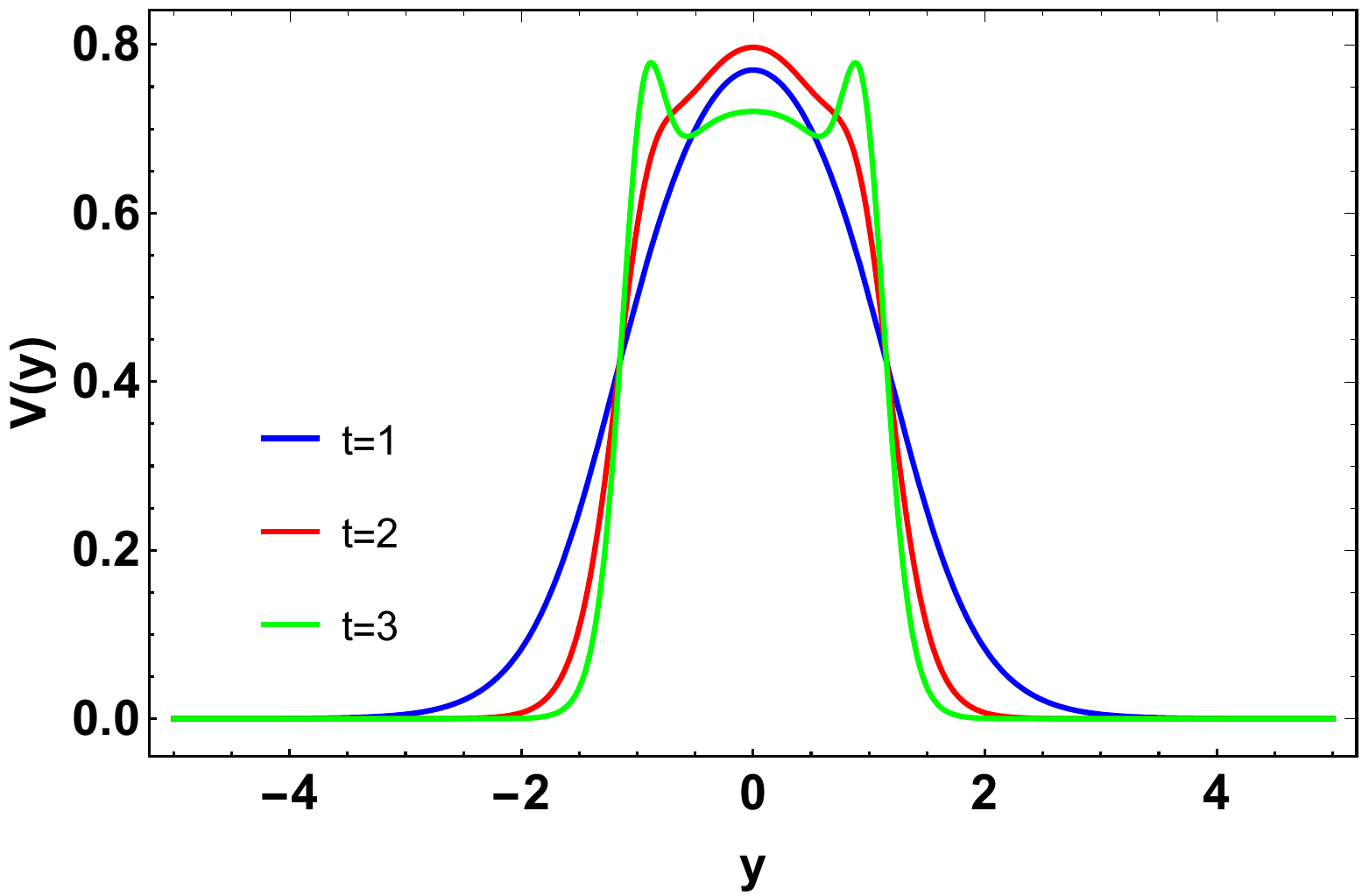}\\
(a)\hspace{6.7cm}(b)\\
\includegraphics[height=5cm]{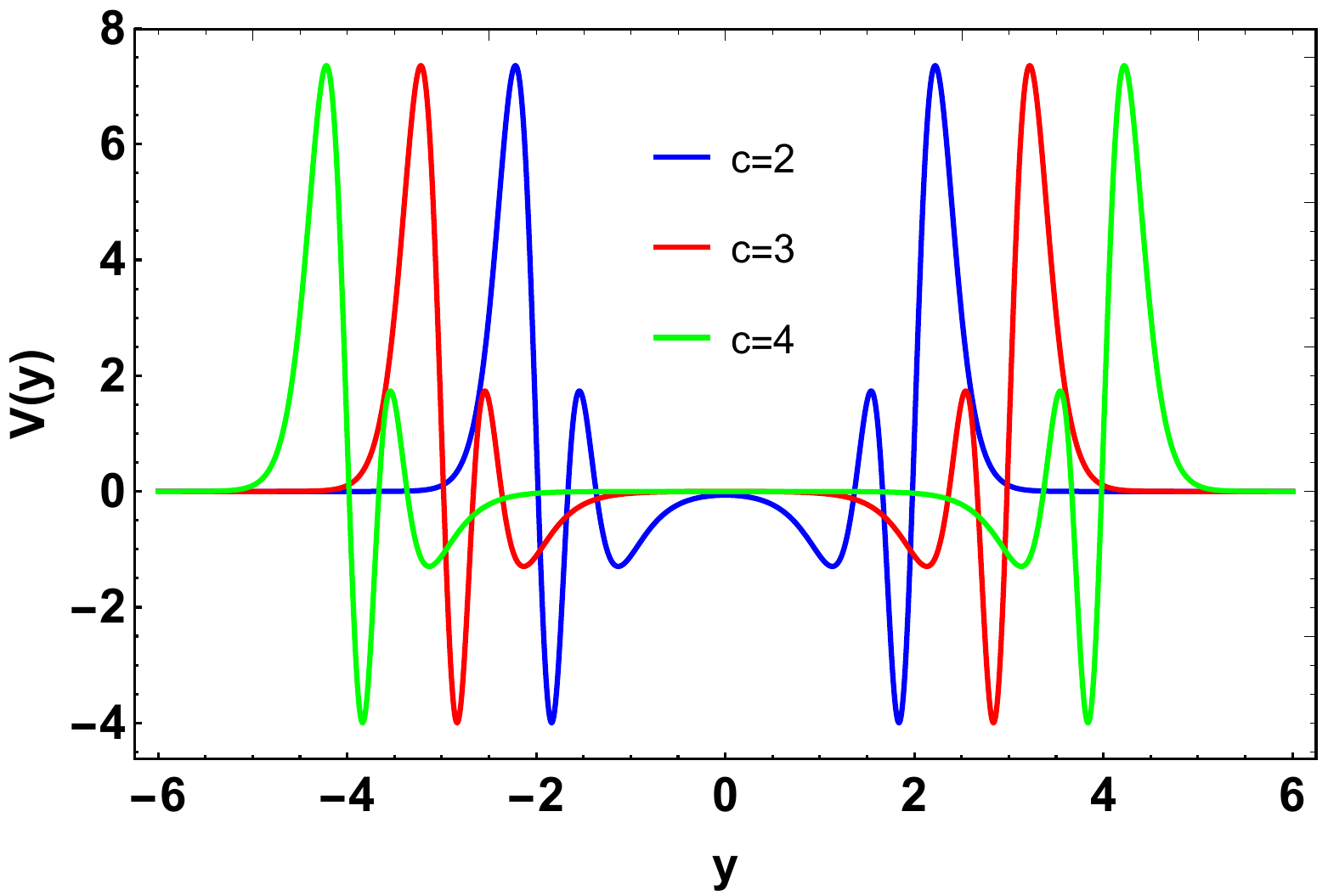} 
\includegraphics[height=5cm]{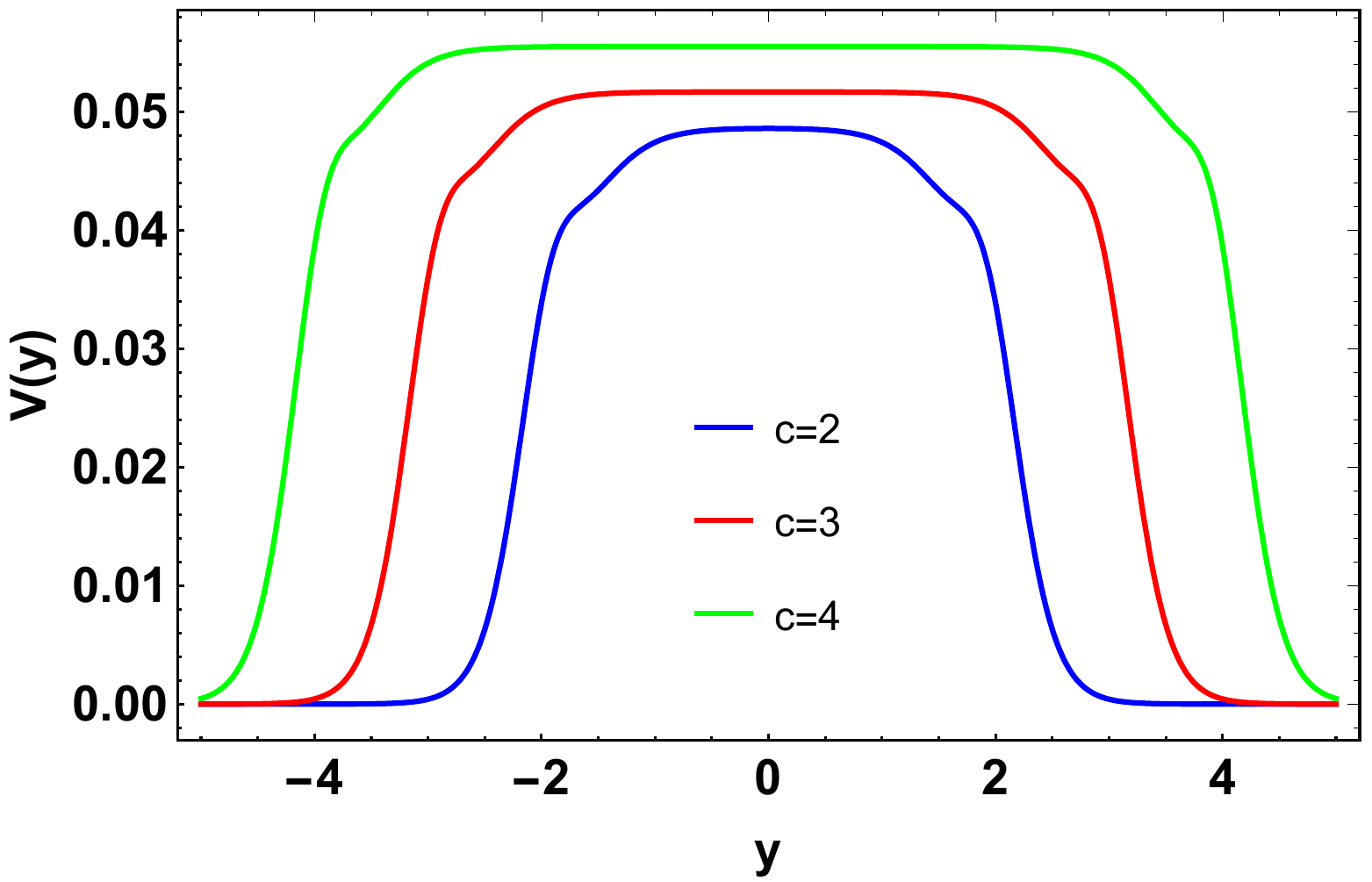}\\
(c)\hspace{6.7cm}(d)
\end{tabular}
\end{center}
\vspace{-0.5cm}
\caption{ For model 2. (a) Effective potential and (b) zero-mode with $t$ varying ($c=1$). (c) Effective potential and  (d) zero-mode with $c$ varying ($t=2$).
\label{fig12}}
\end{figure}

Let us now investigate the behavior of the effective potential and zero-mode for the second model. The shapes of both are depicted in Fig. (\ref{fig12}). The potential shows a volcano-like shape for $t=1$ and a well with an internal structure that tends to a compact profile as we increase the parameter $t$. The zero-mode presents a split for $t=3$, with two peaks separated by a distance controlled by this parameter. Furthermore, we note that the parameter $c$ adjusts the length of the well.

\section{Final remarks}
\label{s5}

In this paper, we have studied a five-dimensional braneworld generated by a single scalar field in nonmetricity-based modified gravity. We have considered different forms of the warp factor to provide a complete description of the braneworld, where $f(Q,\mathcal{T})=Q+k_1Q^n+k_2 \mathcal{T}$, with parameters $k_{1,2}$ related to the influence of nonmetricity and the trace of the energy-momentum tensor, respectively. As demonstrated, the parameters $k_{1,2}$ and $n$ play a significant role in the brane splitting process.

The scalar field solutions exhibit a kink-like profile that can be deformed into a two-kink-like structure by varying the parameters $k_{1,2}$. The energy density is well localized for $n=1$. However, for $n \geq 2$, the energy density shows a phase transition, leading to brane splitting. In Model 2, we have considered a warp factor that exhibits a platform around the brane core. In this case, the matter field solution displays a compact two-kink profile along with the corresponding energy density.

The results obtained for the two warp factor choices, as expected, demonstrate that these models are physically consistent. We proceeded with the study by investigating whether the braneworld scenario in $f(Q,\mathcal{T})$ gravity is stable under linear tensor perturbations. Through gravitational perturbations, we observed the stability of the model. We derived a Schrödinger-like equation, which represents a supersymmetric equation in quantum mechanics, allowing for a normalizable massless mode. As shown, both the effective potentials and the massless modes depend solely on the parameter $k_{1}$, which controls the localization of the zero-mode, making it more or less localized.

As perspectives for future work, we can investigate the localization of massive graviton modes, fermions, and abelian gauge fields in the braneworld studied here. Additionally, it is possible to explore how the massive graviton modes contribute to Newton's Law. Another avenue for investigation is the study of phase transitions in field configurations by employing the configurational entropy (CE) framework.

\section{Acknowledgment}

The authors would like to express grateful to the Funda\c{c}\~{a}o Cearense de Apoio ao Desenvolvimento Cient\'{i}fico e Tecnol\'{o}gico (FUNCAP), the Coordena\c{c}\~{a}o de Aperfei\c{c}oamento de Pessoal de N\'{i}vel Superior (CAPES), and the Conselho Nacional de Desenvolvimento Cient\'{i}fico e Tecnol\'{o}gico (CNPq).  Fernando M. Belchior has been partially supported by CNPq with grant No. 161092/2021-7. Robert V. Maluf thanks the CNPq for grant no. 200879/2022-7. Furthermore, Carlos Alberto S. Almeida is grateful to the CNPq, project number 200387/2023-5.

\end{document}